\documentclass[iop]{emulateapj}
\usepackage{apjfonts}

\begin{document}

\epsscale{1.1}

\title{Flash Mixing on the White Dwarf Cooling Curve: Spectroscopic 
Confirmation in NGC 2808\altaffilmark{1}}
\shorttitle{Flash Mixing on the White Dwarf Cooling Curve}

\author{
Thomas M. Brown\altaffilmark{2}, 
Thierry Lanz\altaffilmark{3,4},
Allen V. Sweigart\altaffilmark{5},
Misty Cracraft\altaffilmark{2}, 
Ivan Hubeny\altaffilmark{6}, and
Wayne B. Landsman\altaffilmark{7}
}

\altaffiltext{1}{Based on observations made with the NASA/ESA {\it Hubble
Space Telescope}, obtained at STScI, which
is operated by AURA, Inc., under NASA contract NAS 5-26555.}

\altaffiltext{2}{Space Telescope Science Institute, 3700 San Martin Drive,
Baltimore, MD 21218;  
tbrown@stsci.edu, 
cracraft@stsci.edu}

\altaffiltext{3}{Department of Astronomy, University of Maryland, College
Park, MD 20742}

\altaffiltext{4}{
Laboratoire Lagrange, UMR7293, Universit\'e 
de Nice Sophia-Antipolis, CNRS, Observatoire de la C\^ote d'Azur, 
F-06304 Nice, France; thierry.lanz@oca.eu}

\altaffiltext{5}{Code 667, NASA Goddard Space Flight Center, Greenbelt, MD
20771; allen.v.sweigart@nasa.gov}

\altaffiltext{6}{Steward Observatory, University of Arizona, Tucson,
  AZ 85712; hubeny@aegis.as.arizona.edu}

\altaffiltext{7}{Adnet Systems, NASA Goddard Space
  Flight Center, Greenbelt, MD 20771; wayne.b.landsman@nasa.gov}

\submitted{To appear in The Astrophysical Journal}

\begin{abstract}

We present new {\it HST} far-UV spectroscopy of two dozen hot evolved
stars in NGC~2808, a massive globular cluster with a large population
of ``blue-hook'' (BHk) stars.  The BHk stars are found in ultraviolet
color-magnitude diagrams of the most massive globular clusters, where
they fall at luminosities immediately below the hot end of the
horizontal branch (HB), in a region of the HR diagram unexplained by
canonical stellar evolution theory.  Using new theoretical
evolutionary and atmospheric models, we have shown that these
subluminous HB stars are very likely the progeny of stars that undergo
extensive internal mixing during a late He-core flash on the white
dwarf cooling curve.  This flash mixing leads to hotter temperatures
and an enormous enhancement of the surface He and C abundances; these
hotter temperatures, together with the decrease in H opacity
shortward of the Lyman limit, make the BHk stars brighter in the extreme
UV while appearing subluminous in the UV and optical.  Our far-UV
spectroscopy demonstrates that, relative to normal HB stars at the
same color, the BHk stars of NGC~2808 are hotter and greatly
enhanced in He and C, thus providing unambiguous evidence of flash
mixing in the subluminous population.  Although the C abundance in the
BHk stars is orders of magnitude larger than that in the normal
HB stars, the atmospheric C abundance in both the BHk and normal
HB stars appears to be affected by gravitational settling.  The
abundance variations seen in Si and the Fe-peak elements also indicate
that atmospheric diffusion is at play in our sample, with all of our
hot subdwarfs at 25,000~K to 50,000~K exhibiting large
enhancements of the iron-peak elements.  The hottest subdwarfs in our
BHk sample may be pulsators, given that they fall in the
temperature range of newly-discovered pulsating subdwarfs in $\omega$ Cen.
In addition to the normal hot HB and BHk stars, we also obtain
spectra of five blue HB stars, a post-HB star, and three unclassified
stars with unusually blue UV colors.

\end{abstract}

\keywords{globular clusters: individual (NGC~2808) -- stars:
  atmospheres -- stars: evolution -- stars: horizontal branch --
  ultraviolet: stars}

\section{Introduction}

For decades, globular clusters have served as the fundamental
laboratory for the study of stellar evolution.  Their utility arose
from the assumption that, within observational errors, each globular
cluster appeared to be comprised of stars with a single age and
chemical composition.  The subsequent discovery that some globular clusters
host multiple stellar generations is one of the most exciting
developments in the study of resolved stellar populations.  These
multiple generations were revealed in the color-magnitude diagrams
(CMDs) of the most massive globular clusters; examples include
the double main sequence (MS) in $\omega$~Cen (Anderson\ 1997) and the
triple MS in NGC~2808 (D'Antona et al.\ 2005; Piotto et al.\ 2007).
The splitting of the MS in these clusters is thought to be due to the presence 
of subpopulations having He abundances as high
as $Y\sim0.4$ (Piotto et al.\ 2005), with these He-rich stars being born
from the He-rich ejecta of the initial stellar generation.

The existence of these He-rich subpopulations
offers the opportunity to test theories of stellar
evolution in a new regime, both in the early evolution on the MS and
in the late stages beyond.
For example, the CMDs of massive clusters exhibit unusual
characteristics on the horizontal branch (HB).  Regardless of
metallicity, massive clusters tend to host significant populations of
extreme HB (EHB) stars at $\rm T_{eff} > 20,000~K$.  These EHB stars
have extremely thin envelopes ($\sim$10$^{-3}$ to $10^{-2}~M_\odot$) -- the result
of extensive mass loss on the red-giant branch (RGB).  The analogs of
the EHB stars in the field are the subdwarf B (sdB) stars, which
produce the ``UV upturn'' in the otherwise cool spectra of elliptical
galaxies (Brown et al.\ 1997; Brown et al.\ 2008).  In general, the
HB of a high-metallicity cluster will be dominated
by red clump stars, while the HB of a low-metallicity cluster will
extend to hotter stars, although other parameters (such as age, He
abundance, and cluster central density) can play a role
in determining the HB morphology
(e.g., Gratton et al.\ 2010; Dotter et al.\ 2010; 
Fusi Pecci \& Bellazzini 1997).  At a fixed cluster
age, the MS turnoff mass decreases strongly with increasing He
abundance, leading to a bluer HB morphology for a given
range of RGB mass loss (D'Antona et al. 2002).  Because massive
clusters are more likely to retain the He-rich ejecta from the
initial burst of star formation, He enrichment may explain the
presence of hot HB stars in those massive clusters that are metal-rich.
For example, in NGC~6388 and NGC~6441, the HB extends to hot temperatures,
and slopes upward in optical CMDs from the red clump to the top of the 
blue HB tail (Rich et al.\ 1997; Busso et al.\ 2007), as one would expect
if these clusters contained a He-rich subpopulation.  
The blue HB morphology of NGC~2808 can be similarly explained by an increasing 
He abundance (from $0.24 < Y < 0.4$) at increasing temperature (Dalessandro
et al.\ 2011).  In this scenario, the EHB stars would be the progeny of the 
most He-rich MS stars.  

Another curiosity in massive clusters is the luminosity dispersion of
their EHB stars.  In seven massive globular clusters hosting EHB
stars, ultraviolet photometry shows that the EHB terminates in a ``blue
hook'' (BHk) of subluminous stars lying up to $\sim$1~mag below the
canonical HB (D'Cruz et al.\ 2000; Brown et al.\ 2001; Brown et
al.\ 2010).  As discussed by Brown et al.\ (2001), 
the most likely explanation for these stars is a delayed
He-core flash.  If the RGB mass loss is large enough,
a star can evolve off the RGB and undergo a delayed He-core
flash either as it crosses the HR diagram (known as an ``early hot flasher'')
or as it descends the white dwarf cooling curve (known as a 
``late hot flasher''; Castellani \& Castellani 1993; D'Cruz et al.\ 1996).
Note that this evolutionary path does {\it not} assume the star evolves in
isolation; although this evolutionary path is frequently described as
``single star evolution'' in the literature, it is in fact driven solely 
by mass loss, whether it occurs in a single star (e.g., stellar winds) or 
in a binary system (e.g., Roche-lobe overflow and mass transfer).
Normally the flash convection
does not penetrate into the envelope, due to the high entropy barrier of
the strong H-burning shell.  However, such penetration is inevitable
if He ignites on the white dwarf cooling curve, where the H-burning shell
is much weaker.  Sweigart (1997) first demonstrated that a flash on the
white dwarf cooling curve will mix the H-rich envelope into the
stellar interior, thereby greatly enhancing the surface He, C and
possibly N abundances.  This result
was subsequently confirmed by detailed calculations
of the flash-mixing phase by Cassisi et al.\ (2003) and
Miller Bertolami et al.\ (2008).  Brown et al.\ (2010)
demonstrated that flash mixing is the only known mechanism that can
plausibly produce the low luminosities of the BHk stars in
massive clusters.  An independent analysis of optical and UV photometry
in NGC~2808 reaffirmed this conclusion (Dalessandro
et al.\ 2011).  The low ultraviolet and optical
luminosities of the flash-mixed stars are primarily due to
their higher effective temperatures (which increases the
bolometric correction) and the reduction in their H opacity
below the Lyman limit (which increases the flux emitted in
the extreme ultraviolet at the expense of the flux at longer wavelengths).

A key prediction of the flash mixing scenario
is a substantial increase in the surface He and C abundances
of the BHk stars.  Spectroscopic evidence in support
of this prediction has already been found in another massive cluster,
$\omega$ Cen, by Moehler et al.\ (2011). 
They obtained optical spectroscopy of potential BHk stars by
selecting targets from the faint end of the hot HB tail in an optical
CMD of the cluster.  They found
that all of the HB stars cooler than 30,000~K were He-poor,
while nearly three-fourths of the hotter stars had solar to
super-solar He abundances, as well as C abundances
up to $\sim$3\% by mass.  Moreover, these C abundances
were strongly correlated with the He abundance.

The He-sdB and He-sdO stars are the most likely field analogs of the
BHk stars found in globular clusters.  Various formation scenarios
have been debated in the literature for this field population,
including both flash mixing and white dwarf mergers (see, e.g., Ahmad
\& Jeffery 2003; Lanz et al.\ 2004; Stroeer et al.\ 2007; Heber \&
Hirsch 2010).  Lanz et al.\ (2004) obtained far-UV spectroscopy of
three He-sdB stars, and demonstrated that the incredibly strong C
lines in two of them (PG1544+488 and JL87) implied an atmospheric
composition of 1--2\% C by mass; they argued that this provided strong
evidence of flash mixing in the field population.  Ahmad et
al.\ (2004) subsequently found that PG1544+488 (the archetype of the
He-sdB class) is a spectroscopic binary of two He-sdB stars; under the
invalid assumption that flash mixing is a ``single star''
evolutionary channel, they argued that binary He-sdB stars present a
problem for the flash mixing scenario.  However, as we noted above, flash
mixing is in fact a natural outcome of extreme mass loss, and does not
depend upon how this mass loss occurs.  Stroeer et
al.\ (2007) noted that the He-sdO stars occupy a narrow temperature
range of 40,000 to 50,000~K, and can be significantly enhanced in C
and/or N.  Heber \& Hirsch (2010) subsequently concluded that the
flash-mixing channel is favored over the merger channel in the
production of those He-sdO stars that are C-rich.  However, the
ongoing debate regarding the origin of the He-sdO and He-sdB stars is
largely due to the uncertain placement of these stars in the HR
diagram relative to the canonical EHB -- specifically, it hinges upon
the accurate measurement of surface gravity and luminosity (cf.\ Ahmad
\& Jeffery 2003; Lanz et al.\ 2004).  Of course, interpretation of an
HR diagram constructed from the field population is hampered by
uncertainties in distance, reddening, age, and original MS chemical
composition.  In contrast, these uncertainties are largely avoided in
a globular cluster population.  Because the BHk populations in the massive
globular clusters observed by Brown et al.\ (2010) are confined to luminosities
immediately below the hot end of the zero-age HB, it seems clear that
white dwarf mergers cannot play a role in the formation of these stars.
If mergers were important, one would expect to find BHk stars with larger
masses and therefore brighter luminosities than the luminosity of the
canonical EHB.  No such stars are observed.  
Most likely, the field population of He-sdB and He-sdO
stars arises from a greater diversity of evolutionary channels,
including possibly mergers, than is the case 
in the globular clusters.

Here, we present recent UV spectroscopy of both normal and subluminous
EHB stars in NGC~2808, where the classification comes from
high-precision UV photometry.  Our objective is to test
the flash-mixing scenario by determining
the He and C abundances in the BHk stars relative to
the abundances in the normal EHB stars; unfortunately, no significant N
diagnostics are available in our spectra.
In total, spectra were obtained for seven normal EHB stars
and eight BHk stars.  The sample of subluminous stars
includes two stars that have luminosities consistent with other BHk
stars but colors significantly redder than the rest of the BHk
population, and indeed far redder than expected from flash mixing.
We also obtained spectra of five blue HB (BHB) stars,
and three unclassified objects with unusually blue UV colors.
Finally, our spectroscopy includes a bright post-HB star too hot to
ascend the asymptotic giant branch (AGB); such stars are usually
classified as AGB-Manqu\'e (AGBM) stars.

\begin{figure}
\plotone{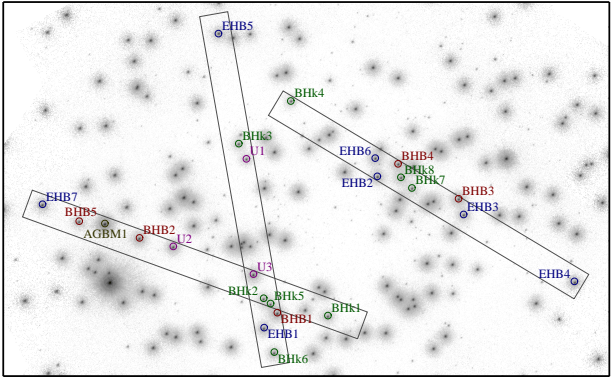}
\caption{A far-UV image of NGC~2808 (Brown et al.\ 2001), shown at a
 logarithmic stretch, with the three spectroscopic slit positions
 indicated (boxes).  Although the slit has dimensions of $52^{\prime\prime}
 \times 2^{\prime\prime}$, the far-UV detector is only $25^{\prime\prime}$ across,
 so we show the slit as truncated by the detector.  The sources with
 clean spectroscopy are labeled according to their
 evolutionary stage: extreme horizontal branch (EHB), blue hook (BHk),
 blue horizontal branch (BHB), AGB-Manqu\'e (AGBM), and
 unclassified (U).  The positions of these stars in the UV CMD of NGC~2808
 are shown in Figure 2.}
\end{figure}

\section{Observations and Data Reduction}

We obtained spatially-resolved spectra
along three slit positions in the center of NGC~2808 using
the Space Telescope Imaging Spectrograph (STIS) on the {\it Hubble
Space Telescope (HST)}.  The program was originally awarded time in 2004,
but was withdrawn before any observations were obtained due to
the failure of a STIS power supply later that year.  The program was
re-proposed and re-awarded time in 2008, but due to scheduling
constraints, the observations were delayed until September 2010 (one
slit position) and February 2011 (two additional slit positions).  The
spectra were obtained with the G140L grating, which provides a
resolution of $\sim$300 km s$^{-1}$, although the spectral purity was
degraded slightly by our use of the wide $52 \times 2^{\prime\prime}$
slit.  We chose this wide slit in order to maximize the number of hot
stars that could be placed within the slit for a given pointing.  The
NGC~2808 core is far less crowded in the far-UV than in the optical,
and so we were able to obtain clean spectra of 24 hot stars using just
three slit positions (Figure 1).  These sources sample various
evolutionary stages in the UV CMD of NGC~2808 (Figure 2).

Each of the three slit positions was observed for 5 orbits, with two
exposures per orbit, giving a total exposure time ranging from 14442
sec to 14460 sec per slit position and a signal-to-noise ratio
(SNR) of $\sim$20 per resolution element.  The resolution and SNR were
intended to discriminate between stars that have and have not
undergone flash mixing, given the enormous differences in atmospheric
abundances between these two possibilities; i.e., the spectra are insufficient
to obtain high-precision abundance estimates.  For a given slit position,
the individual exposures were dithered by a few pixels along the slit,
in order to mitigate detector artifacts and flat-field variations.
Because we obtained spectra of multiple stars per slit position, our
targets are generally offset from the slit midline in the dispersion direction.
To correctly align in wavelength the sensitivity curve and the counts
spectrum of each star, we measured the position within the slit for
each star, both in the dispersion and cross-dispersion directions.
This alignment was an iterative process.  Our initial position
estimates used brief (2 sec) CCD images of the cluster obtained
through the $52^{\prime\prime} \times 2^{\prime\prime}$ slit at the
start of each observing visit, in conjunction with the
far-UV and near-UV images of Brown et al.\ (2001).  We then extracted
the spectra using the IRAF {\sc x1d} package, measured the wavelengths
of strong interstellar lines, and tweaked the position of each star in
the dispersion direction.

\begin{figure}
\plotone{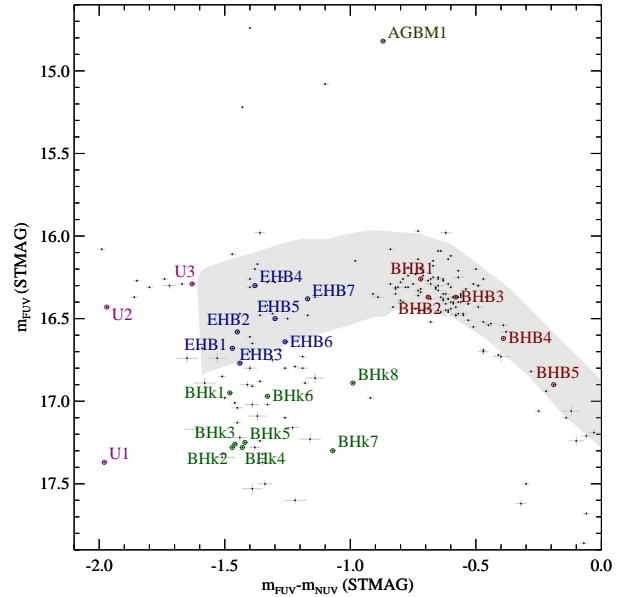}
\caption{The UV CMD of NGC~2808 (Brown et al.\ 2001) with labels for
  those sources with clean far-UV spectroscopy (see Figure 1).  The
  cluster was imaged with the FUV/F25QTZ and NUV/F25CN270 bandpasses
  on STIS (see Figure 2 of Brown et al.\ 2001).  The photometric errors
  are indicated (grey crosses).  The statistical uncertainty in photometric
  color is $<$0.02~mag for our entire spectroscopic sample, and the
  agreement between the observed and theoretical BHB locus indicates
  that systematic errors (e.g., instrument calibration, assumed reddening)
  cannot be large.  The canonical HB locus is shaded grey.}
\end{figure}

By default, the {\sc x1d} package estimates the gross source counts
from an extraction box centered on the object in question, and
subtracts a background estimated from two neighboring extraction
boxes.  To reduce the noise in this background estimate, the
background is smoothed before subtraction.  Specifically, {\sc x1d}
replaces the background counts spectrum with a low-order polynomial
fit to the background, except for two wavelength regions centered on
the bright geocoronal lines of Lyman-$\alpha$ $\lambda$1215 and
\ion{O}{1} $\lambda$1301, where the background spectrum changes too
rapidly to be accurately fit by a low-order polynomial.  These two
geocoronal lines fill the 2$^{\prime\prime}$ slit, such that the lines
are $\sim$50~\AA\ wide in the gross counts spectrum, extending beyond
the wavelength regions normally ignored by the polynomial fit.  To
accurately account for the background in the presence of these broad
geocoronal lines, we turned off background smoothing in {\sc x1d}, and
subtracted the unsmoothed background spectrum at wavelengths shorter than
1350~\AA, while at longer wavelengths, we subtracted a fit to the
background spectrum using a Legendre polynomial.

Although the geocoronal lines are subtracted as part of the background
subtraction in the {\sc x1d} software, the SNR in the net spectrum is
roughly an order of magnitude lower than it is outside of the regions
spanned by these lines.  For a star that is well-centered in the slit,
the geocoronal Lyman-$\alpha$ line will span 1190--1240~\AA\ in the
stellar spectrum, but for a star off-center, the Lyman-$\alpha$ line
can span a region in the wavelength-corrected stellar spectrum that is
offset by $\sim$25~\AA\ in either direction, thus possibly
including a potentially useful \ion{C}{3} multiplet at 1176~\AA.  For
this reason, in our analysis below, the C abundance is generally derived
from both this \ion{C}{3} multiplet and \ion{C}{4}
$\lambda\lambda$1548,1551~\AA.

After finalizing the {\sc x1d} extractions, the individual spectra for
each star were combined with the IRAF {\sc splice} package.  The
combined spectra are shown in Figures 3 -- 10.  For most of the stars,
10 such individual spectra from a single slit position were combined
to produce the final spectrum, but three stars (BHk2, BHk5, and U3) fall
in the overlap between two slit positions (see Figure 1), and thus
have 20 individual spectra and twice the nominal exposure time.  Comparing
the spectra obtained in the two distinct slit positions for these three
stars shows good agreement, providing a check on our extraction procedures.
Besides the 24 hot stars with clean spectroscopy, a handful of other
hot stars fell within the slit for each slit position, but we were
unable to extract accurate spectra for these objects for a variety of
reasons, such as overlapping spectra, spectra falling under the shadow
of the detector repeller wire, spectra falling under a slit occulting
bar, or spectra falling on detector artifacts. 

\begin{figure*}[t]
\plotone{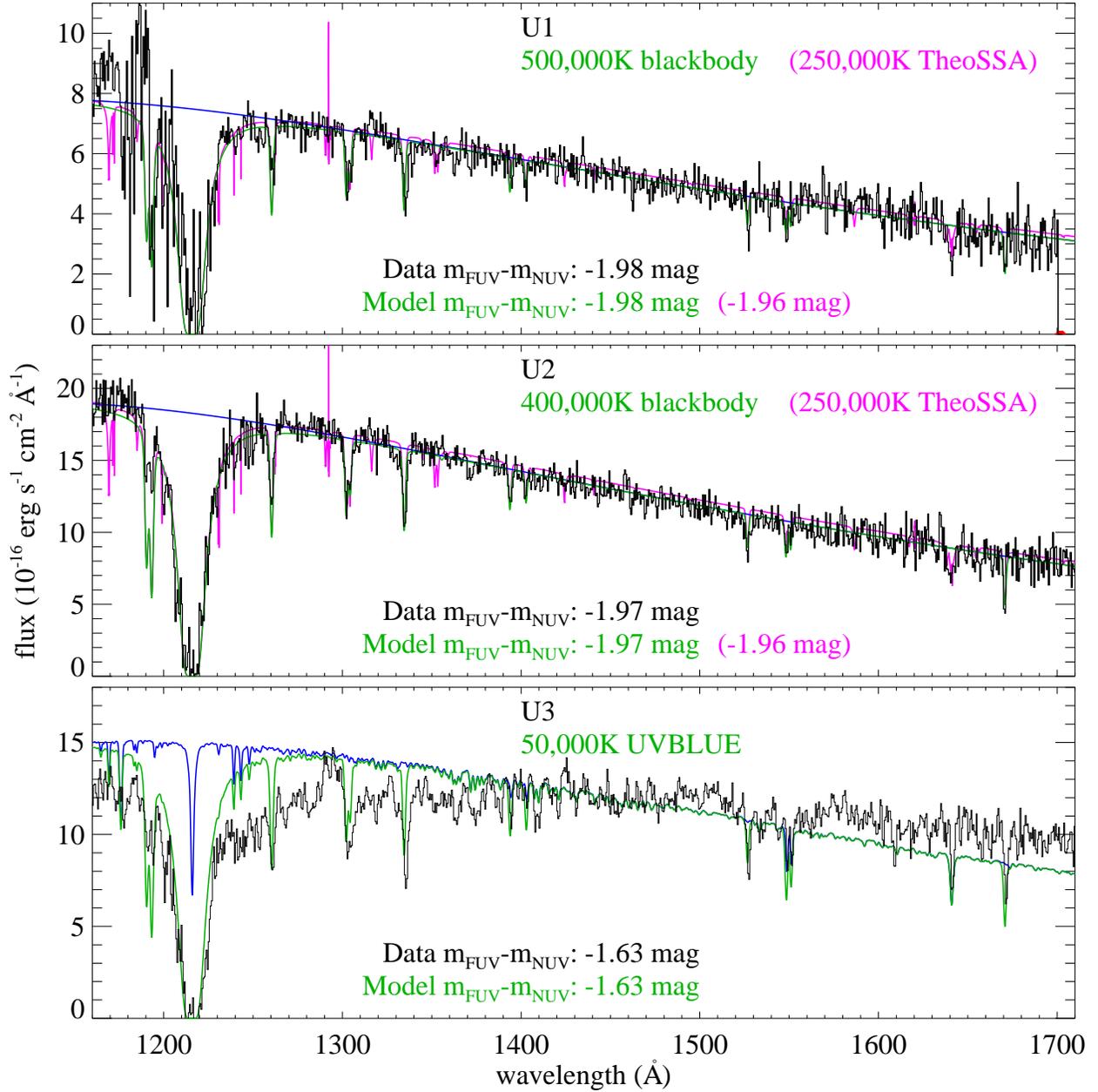}
\caption{The spectra of our 3 unclassified objects (black histograms).
  In the top two panels, we compare the U1 and U2 spectra to
  blackbody models (without ISM absorption, blue; with ISM absorption, green)
  that approximately
  reproduce the far-UV spectral slope and the $m_{FUV}-m_{NUV}$ color,
  although the assumed temperatures are unphysically hot.  We also
  compare these spectra to the hottest synthetic spectrum in the TheoSSA
  database (purple curves; Rauch \& Ringat 2011).  U1 and U2 are nearly
  featureless, other than interstellar absorption features and the
  \ion{He}{2} absorption at 1640~\AA. In the bottom panel, the UV
  photometry of U3 is consistent with a 50,000~K photosphere, but its
  UV spectrum looks much flatter than one would expect for
  this temperature. For comparison, we show the synthetic spectrum of
  a 50,000~K star at [Fe/H]~=$-1.36$ (without ISM absorption, blue; with
  ISM absorption, green) interpolated from the UVBLUE grid 
  (Rodr\'iguez-Merino et al.\ 2005).}
\end{figure*}

\begin{figure*}[t]
\plotone{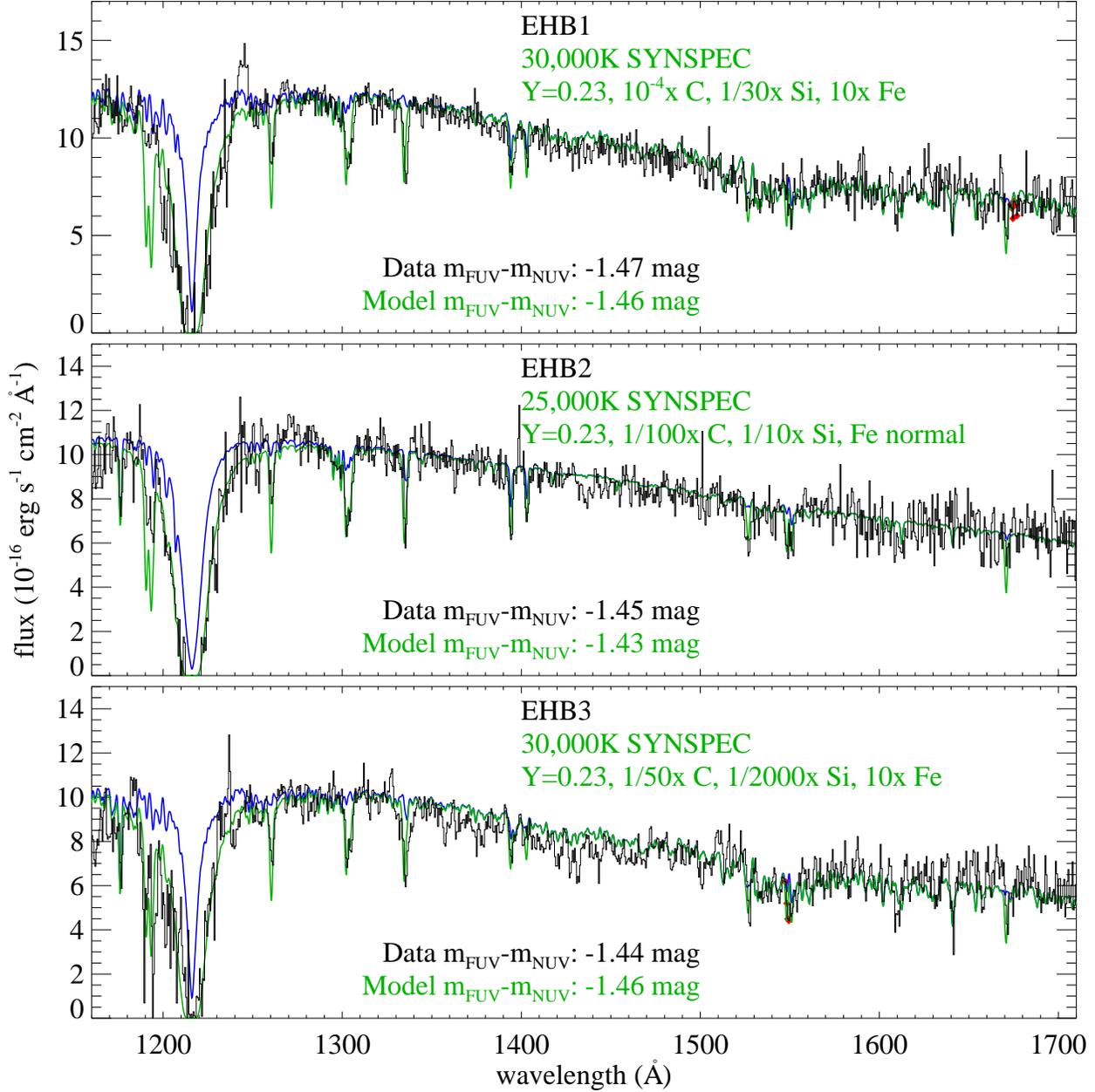}
\caption{The spectra of three normal EHB stars (black histograms)
  compared to SYNSPEC synthetic spectra (without ISM absorption, blue;
  with ISM absorption, green) that approximately match the UV
  photometry and spectroscopy.  The stars exhibit large abundance
  variations relative to the mean cluster abundance at the MS
  (labeled), presumably due to atmospheric diffusion.  Data points
  with potential instrumental artifacts are flagged (red diamonds).  }
\end{figure*}

\begin{figure*}[t]
\plotone{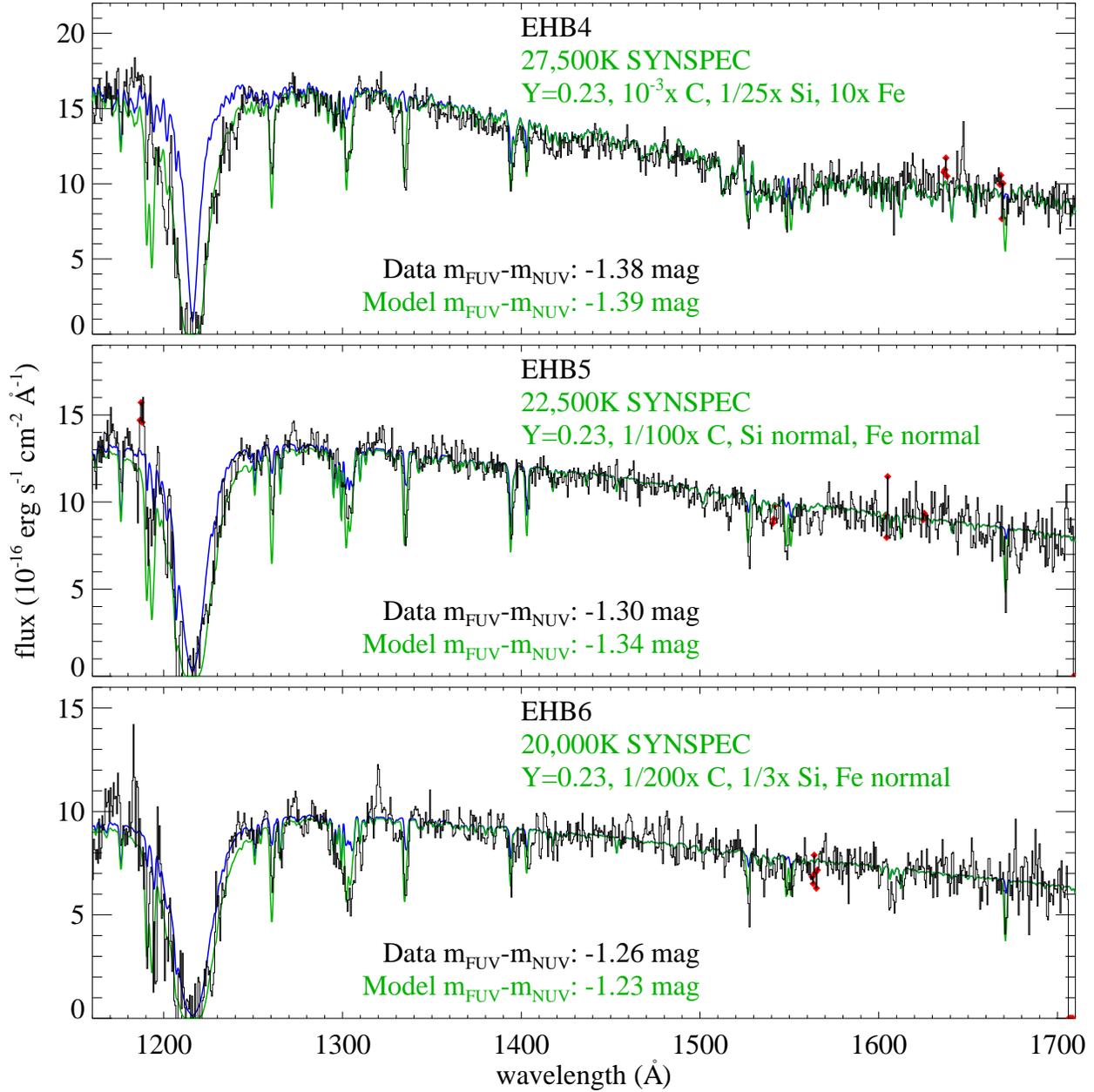}
\caption{As in Figure 4, but for another three EHB stars with normal
  luminosities (black histograms) compared to SYNSPEC synthetic spectra
  (without ISM absorption, blue; with ISM absorption, green).}
\end{figure*}

\begin{figure*}[t]
\plotone{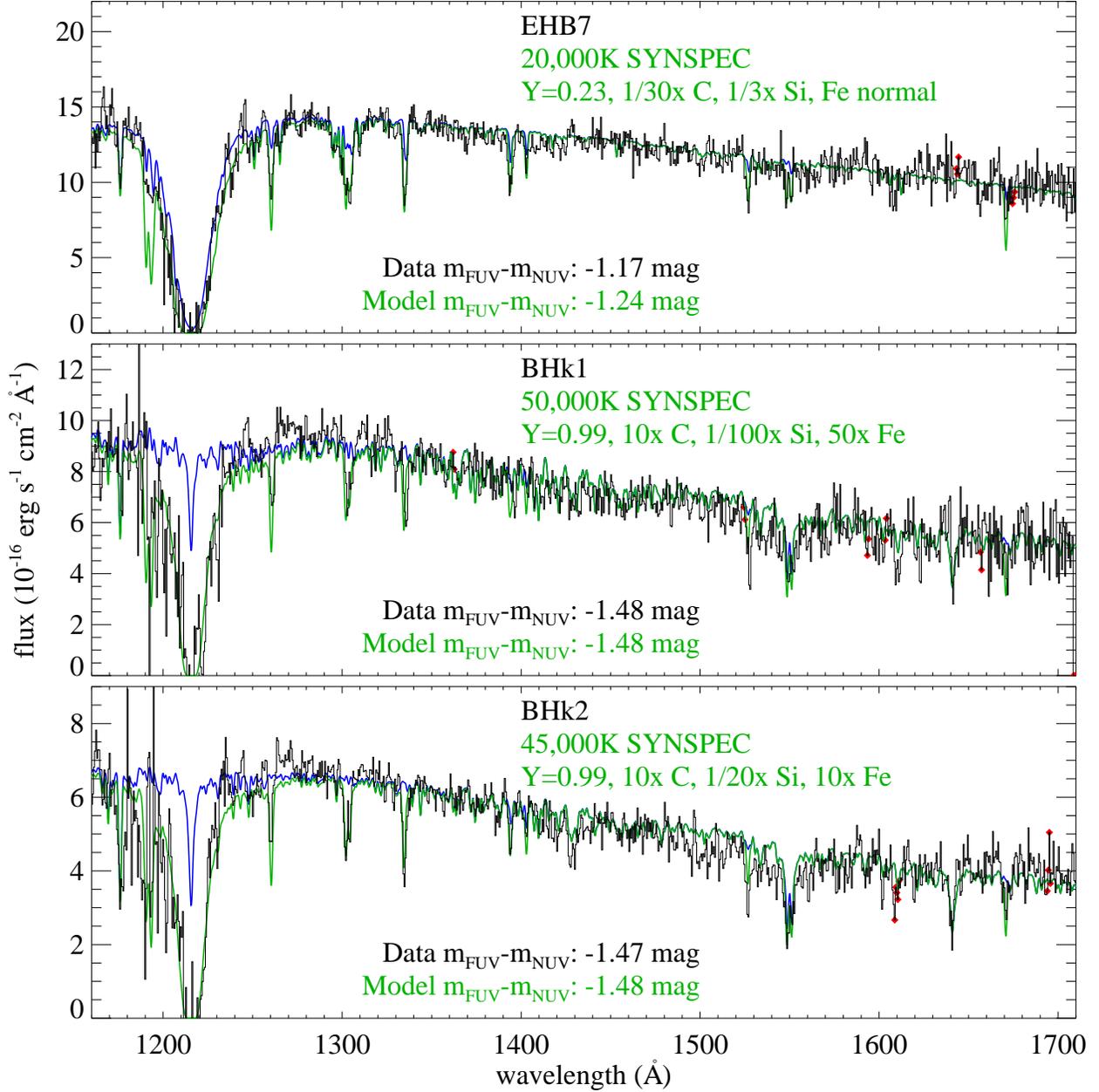}
\caption{As in Figure 4, but for a normal EHB star and two BHk stars
  (black histograms) compared to SYNSPEC synthetic spectra (without
  ISM absorption, blue; with ISM absorption, green).  The BHk stars
  exhibit much stronger He and C lines than the normal EHB stars.}
\end{figure*}

\begin{figure*}[t]
\plotone{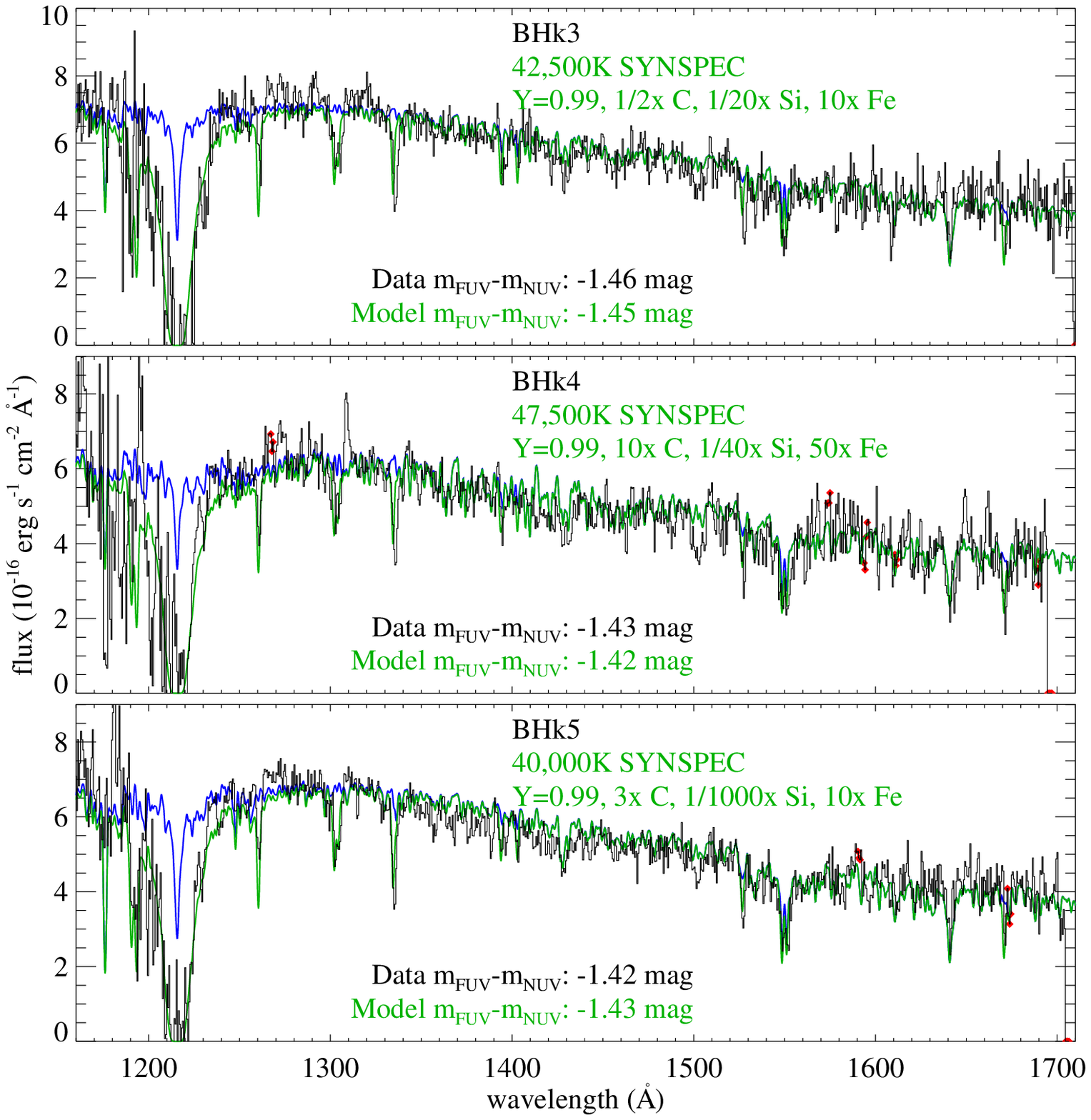}
\caption{As in Figure 4, but for three more BHk stars (black
  histograms) compared to SYNSPEC synthetic spectra (without ISM absorption,
  blue; with ISM absorption, green).}
\end{figure*}

\begin{figure*}[t]
\plotone{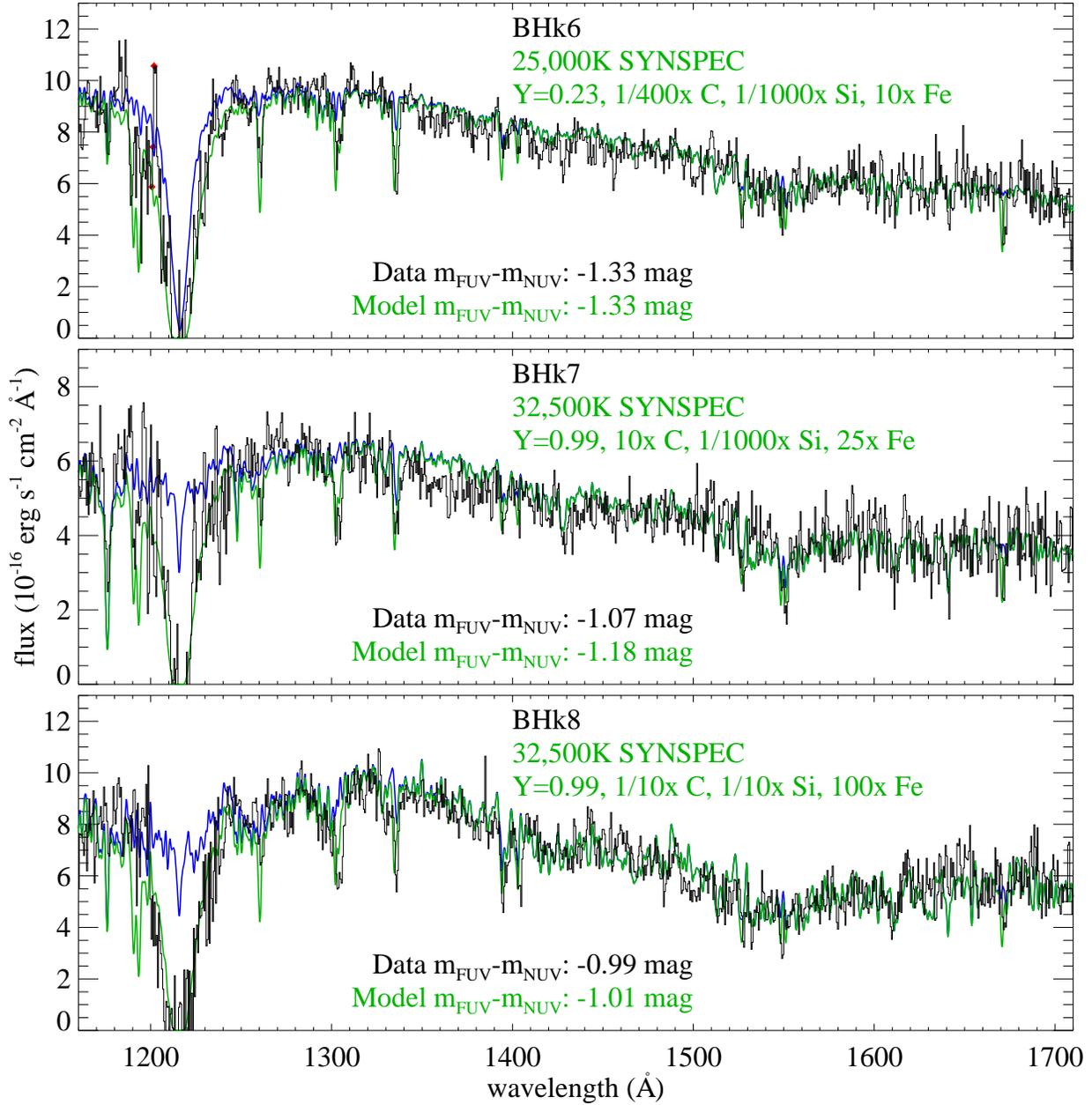}
\caption{As in Figure 4, but for three more BHk stars (black
  histograms) compared to SYNSPEC synthetic spectra (without ISM absorption,
  blue; with ISM absorption, green).
  Unlike the other BHk stars in our spectroscopic sample,
  BHk6 shows no He and C enhancement, is relatively cool, and has a
  luminosity not much below the luminosity of the canonical EHB, so it
  is likely that the star did not undergo flash mixing.  Compared to
  most of the BHk stars in our photometric sample, BHk7 and BHk8 have
  unusually red UV photometry, which may be explained by the strong
  enhancement of Fe, presumably due to atmospheric diffusion. Although
  the BHk8 spectrum does not exhibit a strong \ion{He}{2} feature at
  1640~\AA, an enhanced He abundance in the model is needed to match
  both the UV photometry and far-UV spectroscopic continuum
  simultaneously (see text).}
\end{figure*}

\begin{figure*}[t]
\plotone{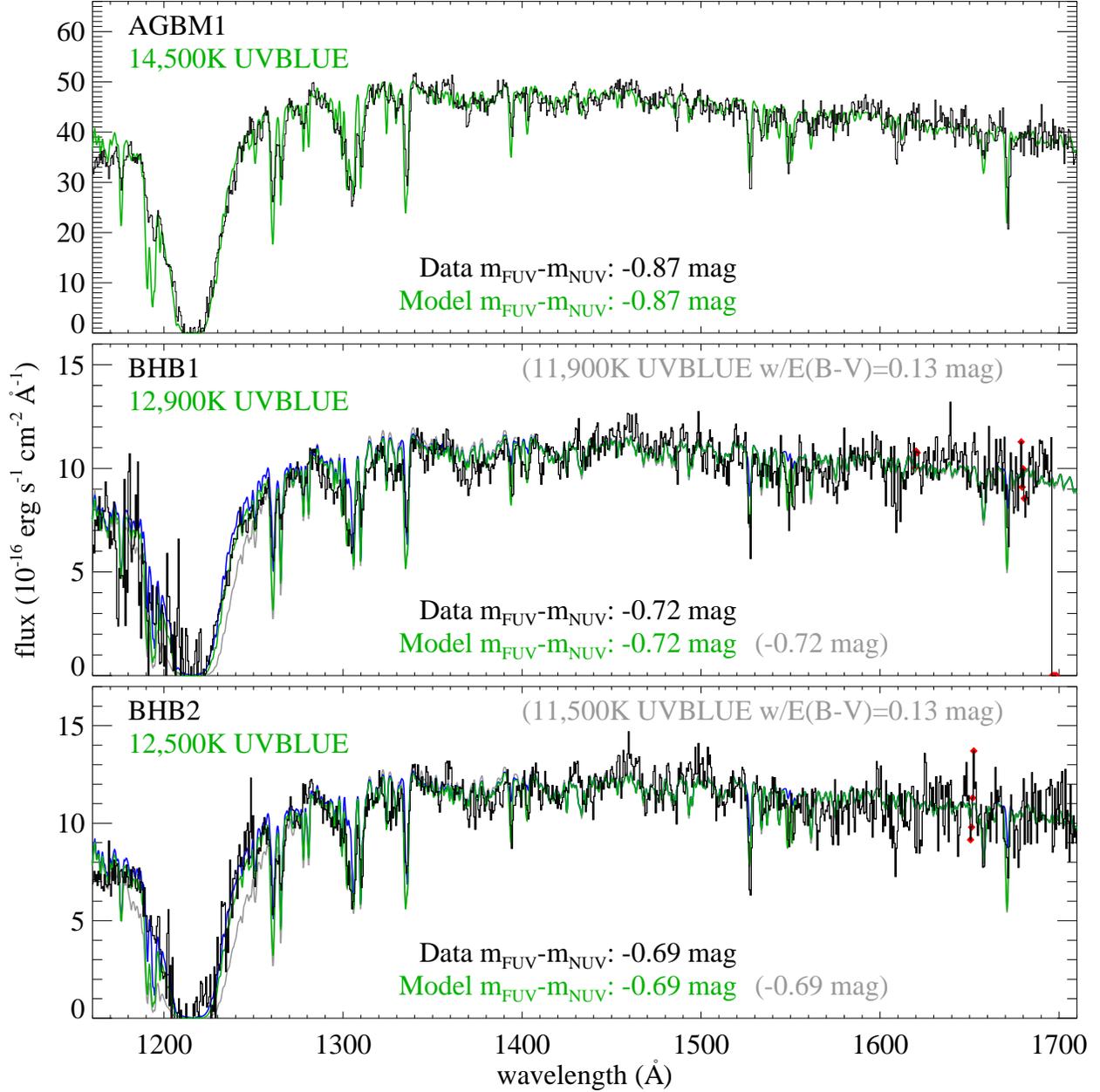}
\caption{As in Figure 4, but for an AGBM star and two normal BHB stars
  (black histograms) compared to UVBLUE synthetic spectra (without ISM
  absorption, blue; with ISM absorption, green).
  The synthetic spectra were interpolated in temperature and
  metallicity from the UVBLUE grid to match the $m_{FUV}-m_{NUV}$
  color at the cluster metallicity.  For BHB1 and BHB2, we also show
  cooler models (grey) with less reddening, which can also 
  reproduce the UV color but not simultaneously the Lyman-$\alpha$ profile.
  Thus, the temperature range of our BHB sample clearly includes the 
  point (at $\sim$12,000~K) where one expects to see a jump in Fe abundance,
  but in fact BHB1 and BHB2 do not exhibit significantly stronger line
  blanketing when compared to the cooler BHB stars in our sample (Figure 10).
}
\end{figure*}

\begin{figure*}[t]
\plotone{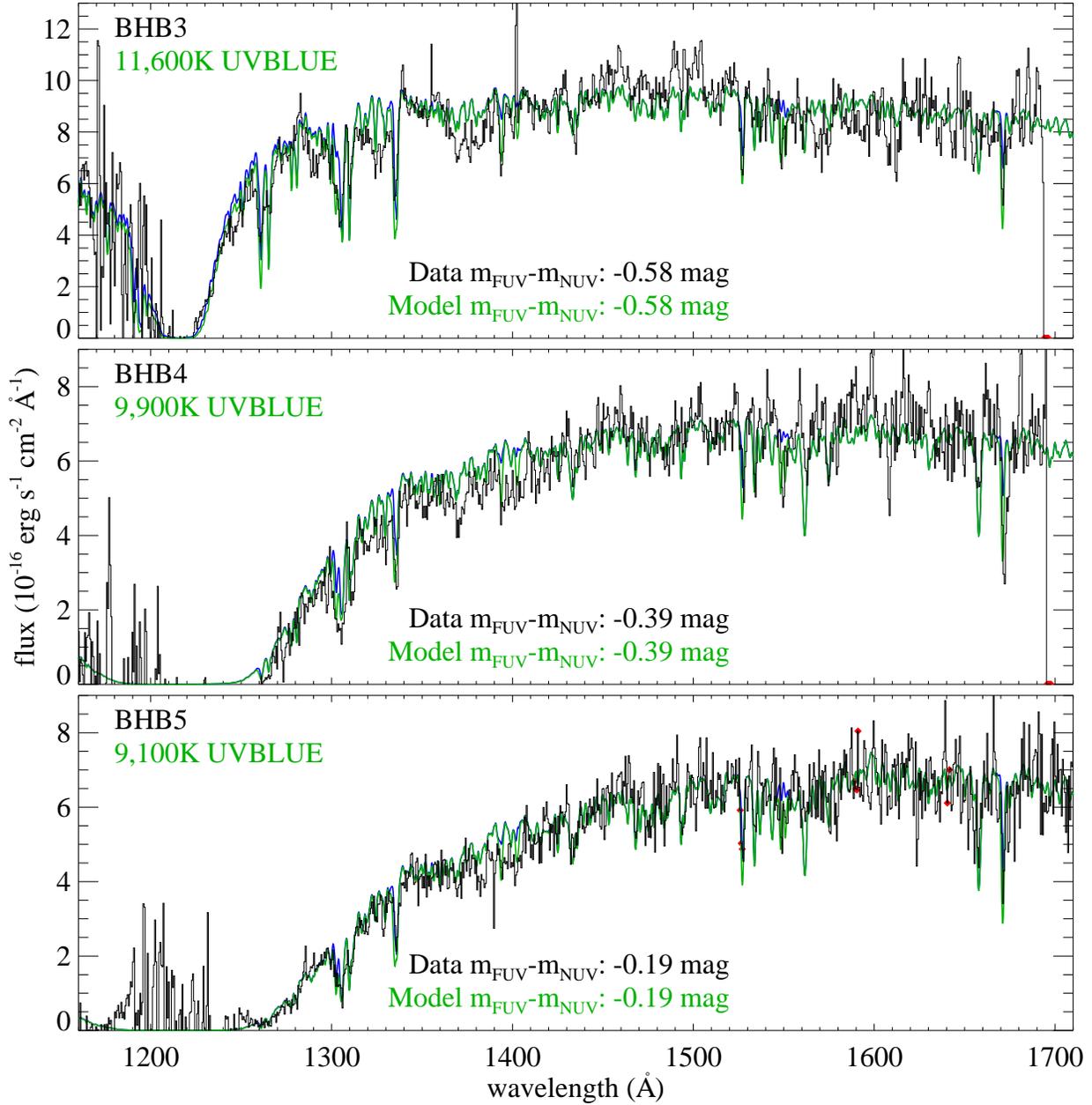}
\caption{As in Figure 4, but for three more normal BHB stars
  (black histograms) compared to UVBLUE synthetic spectra (without ISM
  absorption, blue; with ISM absorption, green).}
\end{figure*}

\begin{figure*}[t]
\plotone{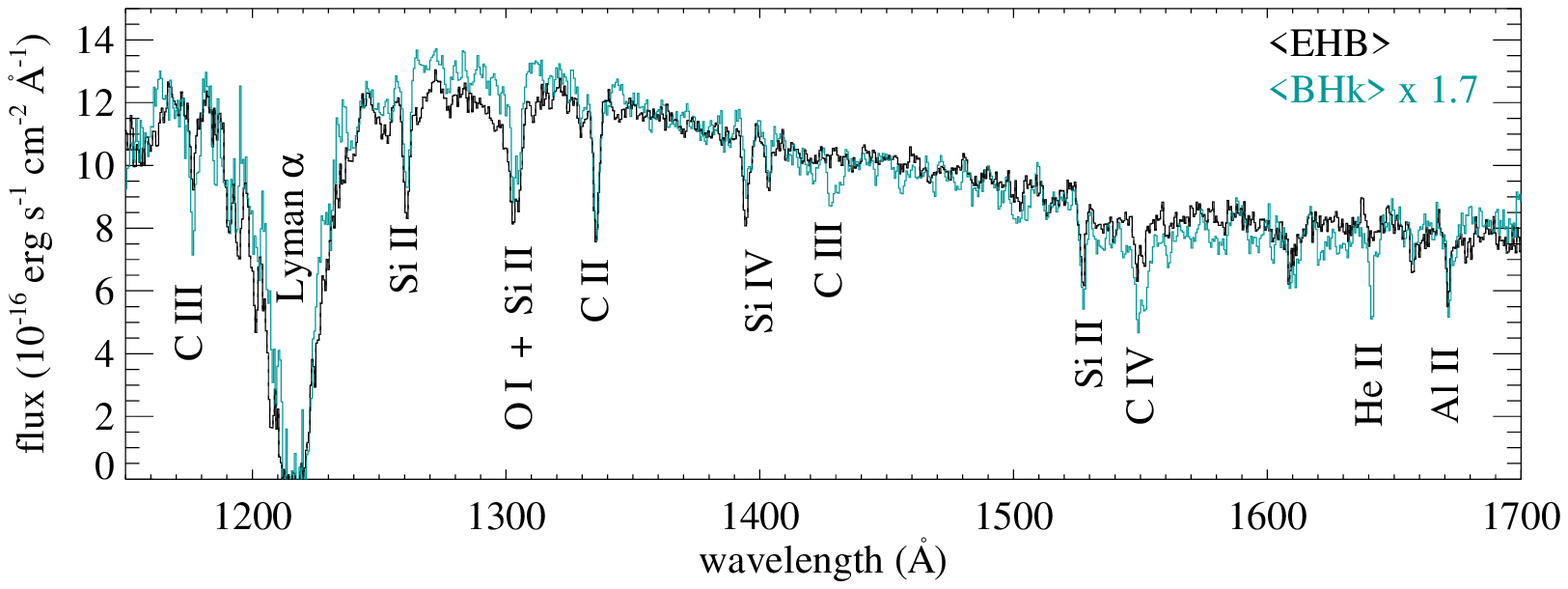}
\caption{The composite spectrum for all EHB stars (black histogram)
compared to that for all BHk stars (cyan histogram).
The BHk spectrum has been multiplied by a factor of 1.7 in order to
normalize the composite luminosity of the BHk stars to that of the EHB stars.}
\end{figure*}

\section{Models}

We interpret our far-UV spectra using synthetic spectra from several
sources.  For the unclassified object U3 and the relatively cool AGBM
and BHB stars, we use the UVBLUE grid (Rodr\'iguez-Merino
et al.\ 2005).  For the hot and nearly featureless spectra of U1 and
U2, we compare to both simple blackbody models and the hottest 
stellar model of Rauch \& Ringat (2011), which has $T_{\rm
  eff}$~=~250,000~K, log~$g$~=7, and mass fractions of 0.33, 0.5,
0.02, and 0.15 for He, C, N, and O, respectively.  For the EHB and BHk
stars that are the primary focus of this paper, we calculated 
non-LTE line-blanketed model atmospheres and synthetic spectra, using
our {\sc Tlusty} (Hubeny \& Lanz 1995) and {\sc Synspec}
programs\footnote{Available at http://nova.astro.umd.edu}.

{\sc Tlusty\/} computes stellar model photospheres in a plane-parallel
geometry, assuming radiative and hydrostatic equilibria.  Departures
from LTE are explicitly allowed for a large set of chemical species
and arbitrarily complex model atoms, using our hybrid Complete
Linearization/Accelerated Lambda Iteration method (Hubeny \& Lanz
1995).  More specifically, the model atmospheres allow for departures
from LTE for 1132 levels and superlevels of 52 ions: \ion{H}{1},
\ion{He}{1}, \ion{He}{2}, \ion{C}{1} -- \ion{C}{4}, \ion{N}{1} --
\ion{N}{5}, \ion{O}{1} -- \ion{O}{6}, \ion{Ne}{1} -- \ion{Ne}{4},
\ion{Mg}{2}, \ion{Al}{2}, \ion{Al}{3}, \ion{Si}{2} -- \ion{Si}{4},
\ion{P}{4}, \ion{P}{5}, \ion{S}{2} -- \ion{S}{6}, \ion{Fe}{2} --
\ion{Fe}{6}.  Details of the model atom setup are provided in Lanz \&
Hubeny (2003, 2007), and in Cunha et al.\ (2006) for updated Ne models.

The model grid spanned 17,500~K to 50,000~K in $T_{\rm eff}$, with
2,500~K steps, and 4.75 to 6.25 in log~$g$, with 0.75 steps.  For the
chemical composition, we initially began with broad abundance
categories.  One grid was calculated at standard cluster abundances,
with [Fe/H]=$-1.36$, $Y=0.23$, and [$\alpha$/Fe]=+0.3.  The other
grids assumed enhanced $Y$ values of 0.4, 0.7, and 0.99, with each
value of $Y$ accompanied by either normal C and N abundances or
enhanced C and N abundances (up to 3\% and 1\% by mass, respectively; see
Lanz et al.\ 2004).  In order to match the absorption line equivalent
widths and broader line-blanketing features in our BHk and EHB spectra,
we generated new models at $Y=0.23$ and $Y=0.99$, with the abundances
of C and Si varied individually, and the abundances of the Fe-peak
elements varied together but independently from the elements outside of
the Fe peak.  In these models, we assumed log~$g$=5.5, which should be
representative of the surface gravities in the BHk and EHB stars, given
the insignificant variations in the low-resolution far-UV spectral
features over the full range of surface gravity in such stars.

After the detailed emergent UV spectrum for each model
atmosphere was calculated with {\sc Synspec}, they were shifted
to the radial velocity of NGC~2808 (101.6 km s$^{-1}$; Harris 1996).
We then added the absorption from strong interstellar lines of 
\ion{H}{1}  (1.9$\times$10$^{21}$ cm$^{-2}$), 
\ion{C}{2}  (6.8$\times$10$^{18}$ cm$^{-2}$),
\ion{C}{4}  (6.0$\times$10$^{17}$ cm$^{-2}$), 
\ion{Si}{2} (1.7$\times$10$^{17}$ cm$^{-2}$),
\ion{Si}{4} (7.6$\times$10$^{16}$ cm$^{-2}$),
\ion{O}{1}  (6.0$\times$10$^{18}$ cm$^{-2}$), and
\ion{Al}{2} (5.0$\times$10$^{16}$ cm$^{-2}$),
using the observed spectra of U1 and U2 as a guide
(Figure 3), given the lack of obvious photospheric features in these
spectra (aside from \ion{He}{2}).  Our assumed \ion{H}{1}
column is nearly twice what one would expect from the mean gas-to-dust
ratio in the Galaxy (e.g., Bohlin et al.\ 1978), but there are significant
variations in this ratio along any given sightline (e.g., Diplas \&
Savage 1994).  It is also larger than the column of
1.2$\times$10$^{21}$ cm$^{-2}$ found in the x-ray analysis of Servillat
et al.\ (2008), but a column that low is strongly discrepant with 
the Lyman-$\alpha$ profile in our spectra.  
We then reddened the spectrum using the mean Galactic
extinction curve of Fitzpatrick (1999), assuming $E(B-V)=0.18$~mag (Brown et 
al.\ 2001, 2010). 

\section{Analysis}

\subsection{Comparison of the Composite EHB and BHk Spectra}

We begin our analysis with an empirical look at the composite spectra of both
the EHB and the BHk samples.  The stars in each sample span a similar range of
$m_{FUV}-m_{NUV}$ color, but the BHk stars are $\sim$0.7~mag fainter than the
EHB stars (Figure 2).  In Figure 11, we show the composite spectrum for
all 7 of the normal EHB stars in our sample, compared to the composite
spectrum for all 8 of the BHk stars (i.e., subluminous EHB stars) in
our sample.  Several strong features are due almost exclusively to
interstellar absorption, and are similar in both of the
composite spectra: \ion{Si}{2} $\lambda$1260~\AA, \ion{O}{1}
$\lambda$1301~\AA, \ion{Si}{2} $\lambda$1304~\AA, \ion{C}{2}
$\lambda$1335~\AA, \ion{Si}{2} $\lambda$1527~\AA, and \ion{Al}{2}
$\lambda$1670~\AA.  At these temperatures, the \ion{Si}{4}
$\lambda\lambda$1394,1403~\AA\ doublet has significant contributions from
both photospheric and interstellar absorption, but the strength of
this feature is similar in both composite spectra, implying that on
average, there is not much difference in Si abundance between the two
populations.  The Lyman-$\alpha$ feature in these spectra is dominated
by interstellar absorption, but the BHk composite spectrum clearly
exhibits less Lyman-$\alpha$ absorption.  Given the dominant
contribution from interstellar absorption in this feature, the
photospheric absorption must be much weaker in the BHk stars than in
the EHB stars, in order to produce a noticeable difference in the
combined interstellar and photospheric feature.  The presence of
weaker Lyman-$\alpha$ absorption in the BHk stars is consistent with
both higher temperatures and/or a higher He abundance in the
atmospheres of the BHk stars, which is what one would expect if the BHk
stars are flash-mixed.  Besides this difference in Lyman-$\alpha$,
there are strong distinctions between the BHk and EHB stars in 
four other absorption features: \ion{C}{3} $\lambda$1176~\AA\ (a multiplet
of 6 lines), \ion{C}{3} $\sim$1427~\AA\ (from 15 lines spanning 1424--1429~\AA),
\ion{C}{4} $\lambda\lambda$1548,1551~\AA, and \ion{He}{2}
$\lambda$1640~\AA\ (a triplet).
The \ion{C}{3} and \ion{He}{2} features
are purely photospheric, while the \ion{C}{4} feature includes both
interstellar and photospheric contributions.  The C and He features
are clearly stronger in the composite BHk spectrum, as expected if the
BHk stars are flash-mixed.  These features become stronger at the higher
abundances and higher temperatures expected in flash-mixed stars
(Brown et al.\ 2001).

\subsection{Individual EHB and BHk Spectra}

We next turn to the individual spectra of the EHB and BHk stars in our
sample.  As stated earlier, our synthetic spectra were computed on a $T_{\rm
 eff}$ grid with 2,500~K spacing, initially employing broad
abundance classes.  For each star, we first selected a synthetic
spectrum that reproduced the gross characteristics of the observed
spectrum and photometry, and then altered the abundances of C, Si, and
the Fe-peak elements to match the observed atmospheric features.  At
the resolution and SNR of our spectra, estimating the equivalent width
is hampered by the determination of the reference pseudo-continuum
level in the presence of so many other lines.  However, the data are
of sufficient quality to characterize gross distinctions in abundance,
at the level needed to detect the signature of flash mixing.  

For simplicity, we will specify the abundances of C, Si, and the
Fe-peak elements as multipliers on the abundances one would expect
from a population at [Fe/H]~=~$-1.36$ with [$\alpha$/Fe]~=~0.3.
However, it is worth repeating that NGC~2808 hosts a triple main
sequence (D'Antona et al.\ 2005; Piotto et al.\ 2007), with the
reddest and bluest sequences exhibiting clear chemical distinctions
(Bragaglia et al.\ 2010). In particular, Bragaglia et al.\ (2010)
found that a star on the blue MS in NGC~2808 had a lower C abundance
([C/Fe]~=$-0.7$) than a star on the red MS ([C/Fe]~=$-0.4$), as one
would expect if the blue MS stars in NGC~2808 were formed from the
He-rich ejecta of the first stellar generation.  The C abundance in
these MS stars would decrease even further due to the envelope mixing
known to occur in globular cluster stars during the evolution up the
RGB (Kraft 1994).  Thus, one would expect the C abundance in the EHB
stars of NGC~2808 to be strongly depleted, especially if they are the
progeny of the blue MS stars.  In addition, atmospheric diffusion
might decrease the C abundance in these EHB stars even further (Miller
Bertolami et al.\ 2008).

Although the composite spectra of the BHk and EHB samples in Figure 11
show no obvious systematic difference in the Si abundance between the
two classes, the Si abundance does show large star-to-star variations
within each group (Figures 3--8).  Most of the detectable Si features
in our far-UV spectra are dominated by interstellar absorption, and none
are completely photospheric, but \ion{Si}{4}
$\lambda\lambda$1394,1403~\AA\ does provide a rough indication of the
stellar Si abundance.  One star (EHB5; Figure 5) exhibits a Si
abundance close to the cluster value, but most of the EHB and BHk stars
exhibit weak to strong depletion of Si relative to the cluster value.
For other stars (e.g., BHk5, BHk6, and BHk7; Figures 7 \& 8), Si is
depleted by at least a factor of 1000.  For photospheric Si abundances
that are orders of magnitude below the cluster value, the \ion{Si}{4}
feature is almost completely dominated by interstellar absorption, and
the photospheric abundance we measure is an approximate upper limit.
In general, our Si abundance is only accurate
to an order of magnitude (see Figure 12).

With these spectra, there are no sufficiently isolated absorption
lines from the Fe-peak elements that can be used to accurately
characterize their abundances.  However, the Fe-peak elements
do cause broad absorption troughs in the far-UV spectrum that require
large changes in the Fe-peak abundances in order to match the observed
variations in the pseudo-continuum (Figure 13).  
The most obvious examples of this
absorption can be seen in BHk7 and BHk8 (Figure 8), where the Fe-peak
abundances are enhanced over the cluster value by factors of 25 and
100, respectively.  Brown et al.\ (2001, 2010) have noted that some of
the BHk stars in massive globular clusters are curiously red -- much
redder than one would expect from models for either normal EHB stars
or BHk stars.  BHk7 and BHk8 are in this group of curiously red BHk stars.
Our spectra for these BHk stars show that their red colors can be
explained by a large enhancement in their abundances of the Fe-peak
elements.  For comparison, models at the mean cluster abundance
can match the $m_{FUV}-m_{NUV}$ colors of BHk7 and BHk8 
only at much cooler temperatures of 16,900~K and 15,600~K,
respectively; however, at such temperatures, the shape of the far-UV
continuum in the synthetic spectra looks nothing like that observed.
These large Fe-peak enhancements are required for the model to
reproduce both the UV photometry and spectroscopy simultaneously.  All
of the BHk stars show Fe-peak enhancements, ranging from 10$\times$ to
100$\times$ the cluster value, while the EHB stars are mixed, with
three showing a 10$\times$ enhancement and the other four showing no
evidence for enhancement.  Because the enhancement of Fe-peak elements
does not come from the characterization of individual lines, 
the abundance uncertainty is approximately a factor of $\sim$3 (Figure 13).

There are three clean C features in the far-UV spectra: the
purely photospheric feature at 1176~\AA\ from the \ion{C}{3}
multiplet, the purely interstellar feature at 1335~\AA\ from the
\ion{C}{2} multiplet, and the \ion{C}{4}
$\lambda\lambda$1548,1551~\AA\ doublet that has both photospheric and
interstellar contributions.  There is also a photospheric
feature at $\sim$1427~\AA\ arising from 15 \ion{C}{3} lines (see
Figure 11), but it is difficult to characterize in individual spectra,
given its strength and blending with other features. Although the
\ion{C}{3} 1176~\AA\ feature potentially provides the best indicator of the
stellar C abundance, in practice it frequently suffers from low SNR,
relative to the \ion{C}{4} feature.  This is because the \ion{C}{3} 
1176~\AA\ feature is affected to a varying degree by the Lyman-$\alpha$
geocoronal line (see \S2). For this reason, we try to match the
strength of both the \ion{C}{3} 1176~\AA\ and \ion{C}{4} 1550~\AA\ features.  
In all of the EHB stars, the C abundance is depressed relative to that in the
cluster, to varying degrees: C is depressed by a factor of 30 in EHB7,
whereas C is nearly undetectable in EHB1, where it is depressed
by a factor of 10,000.
In the BHk sample, five of the stars (BHk1, BHk2, BHk4,
BHk5, and BHk7) are enhanced in C (by factors of 3 to 10), two of the
stars (BHk3 and BHk8) are somewhat depressed in C (by factors of 2 to
10), and one star (BHk6) is significantly depressed in C (by a factor
of 400).  Given the SNR at \ion{C}{3} and the interstellar contamination
of \ion{C}{4}, our C abundances are generally accurate to a factor of $\sim$2 
(see Figure 12), but for those stars where the C abundance is orders
of magnitude below the cluster value, our measured abundance is an
approximate upper limit.

There is one significant He feature in the far-UV spectra: the purely
photospheric \ion{He}{2} feature at 1640~\AA.  At the temperatures of
EHB and BHk stars, the feature is mildly sensitive to abundance and
very sensitive to $T_{\rm eff}$.  Furthermore, for these temperatures,
the $m_{FUV}-m_{NUV}$ color does not significantly change if the
He abundance is increased from $Y=0.23$ to $Y=0.4$, but it becomes
$\sim$0.1--0.3~mag redder if the He is increased to $Y=0.99$.  For
this reason, the He abundance must be constrained by matching the
far-UV spectral slope, the $m_{FUV}-m_{NUV}$ color, and the
\ion{He}{2} absorption feature simultaneously.  In the EHB sample,
these three aspects of each star can be approximated by a model at the
standard cluster He abundance ($Y=0.23$).  
The \ion{He}{2} feature is difficult to characterize in the EHB sample, 
given the weakness of the feature at these cooler temperature, 
the relatively low SNR at the red end of the
STIS spectrum, and the line blanketing in the vicinity of the feature;
it is only obvious in the hottest stars of our EHB sample (EHB1 and EHB3).
Although He is often depleted in 
EHB stars (e.g., see Moehler et al.\ 2011), reducing the
He abundance from $Y=0.23$ to $Y=0.01$ has no significant effect
on the far-UV spectral slope, while the $m_{FUV}-m_{NUV}$ color only changes
at the $\sim$0.01~mag level.  That said, in those EHB stars where the feature
is seen (EHB1 and EHB3), the equivalent width of \ion{He}{2}
is closer to that in a $Y=0.23$ model than a $Y=0.01$ model (Figure 14);
for the remaining EHB stars, the He abundance could be significantly
depleted, and we would have no way of measuring it from the UV spectrum or 
photometry.  In contrast, the \ion{He}{2} feature is extremely strong in six 
of the BHk stars (BHk1, BHk2, BHk3, BHk4, BHk5, and BHk7), even though 
these stars have $m_{FUV}-m_{NUV}$ colors that are similar to those in the EHB 
sample.  To simultaneously match the $m_{FUV}-m_{NUV}$ color and \ion{He}{2}
feature in each of these stars, the synthetic spectra must be much
hotter than those used to match the EHB sample, with an atmosphere
that is 99\% He by mass (Figure 14).  

To see why this is the case, we can compare stars in the EHB and BHk
samples that have similar $m_{FUV}-m_{NUV}$ colors.  For example, EHB1
(Figure 4) and BHk2 (Figure 6) have exactly the same $m_{FUV}-m_{NUV}$
color ($-1.47$~mag), they both exhibit enhanced abundances of the Fe
peak elements (10 times larger than the cluster mean), and they have
similar Si abundances (20 to 30 times lower than the cluster
mean).  However, the BHk2 spectrum has much stronger \ion{C}{4} and
\ion{He}{2} lines than the spectrum of EHB1.  The EHB1 spectrum is
well-matched by a model at 30,000~K, but the BHk2 spectrum requires a
model that is 15,000~K hotter.  If one takes the BHk2 model shown in
Figure 6 and reduces $Y$ from 0.99 to 0.23 (while holding all other
parameters fixed), the $m_{FUV}-m_{NUV}$ color increases to
$-1.60$~mag, which is far bluer than the $-1.47 \pm 0.014$~mag
observed.  Only a He-rich model reproduces the spectral slope,
\ion{He}{2} feature, and $m_{FUV}-m_{NUV}$ color.  A similar argument
can be made comparing EHB4 (Figure 5; $T_{\rm eff}$=27,000~K,
$m_{FUV}-m_{NUV}$=$-1.38$, 10$\times$ enhanced Fe-peak elements,
25$\times$ depleted Si) and BHk5 (Figure 7; $T_{\rm
  eff}$=40,000~K, $m_{FUV}-m_{NUV}$=$-1.42$, 10$\times$ enhanced
Fe-peak elements, 1000$\times$ depleted Si); EHB4 has very weak
\ion{C}{4} absorption and no detectable \ion{He}{2} absorption, while
BHk5 has extremely strong \ion{C}{4} and \ion{He}{2} absorption.

While most of the BHk sample exhibits obvious He enhancement, there
are exceptions.  The photometry and spectroscopy of BHk6 can be matched
by a model that has a relatively low $T_{\rm eff}$ (25,000~K),
depleted C abundance (400$\times$ lower than the cluster value), and
$Y=0.23$.  The BHk8 spectrum does not appear to have a strong
\ion{He}{2} feature, but a $Y=0.99$ model is required to
simultaneously match the photometry and spectroscopic slope; reducing
$Y$ to 0.23, while holding the other parameters fixed, does not
significantly impact the agreement with the far-UV spectrum, but makes
the model $m_{FUV}-m_{NUV}$ color 0.3~mag bluer.  Similar to BHk6, the
BHk8 spectrum indicates a C abundance depleted with respect to the
cluster value, although it is still much higher than that in the EHB
sample.

To summarize, the EHB stars and BHk stars exhibit no systematic
difference in Si abundance, but both classes exhibit large
star-to-star variations in Si abundance, probably resulting from
atmospheric diffusion.  The Fe abundance also varies strongly from
star to star, but it is systematically higher in the BHk stars than in
the EHB stars, and all of the BHk stars exhibit large Fe enhancements.
The largest chemical distinctions between the EHB and BHk stars are
those related to flash mixing (He and C).  All of the EHB stars
exhibit C abundances much lower than the cluster value, and He
abundances at or below the solar value.  As a group, the BHk stars are
significantly hotter than the EHB stars, with five of them exhibiting
enhanced C, and seven of them exhibiting enhanced He, which is strong
evidence that most of the BHk population arises from flash mixing.  The
fact that He and C are not enhanced in the full BHk sample may indicate
that some of these stars did not undergo flash mixing (e.g., BHk6), or
that these elements were depleted due to atmospheric diffusion (e.g.,
BHk3 and BHk8; see Miller Bertolami et al. 2008). 

\subsection{Hot Unclassified Stars}

Although the focus of this project is a comparison of the EHB and BHk
samples in NGC~2808, our spectroscopic sample includes 9 other hot
stars, which we briefly discuss here.  Three of these stars (U1, U2,
and U3) are hotter than the canonical HB (see Figure 2), and are
labeled as unclassified.  As with our BHk and normal EHB samples, the
photometric uncertainties on these stars are small (see Figure 2), and
there does not appear to be anything unusual about them in the UV
images of the cluster (e.g., they do not appear to be blends, or to
suffer from a detector artifact).  Both U1 and U2 are hotter (bluer)
than the evolutionary path a low-mass star would take through the HR
diagram as it begins to descend the white dwarf cooling curve, and
they are also significantly brighter than the part of the white dwarf
cooling curve where stars would begin to appear in any significant
numbers, given the speed of the evolution between the HB and the white
dwarf phases (e.g., see Figure 3 of Brown et al.\ 2001).  The
$m_{FUV}-m_{NUV}$ colors of U1 and U2 are implausibly blue compared to
the expectations from evolutionary tracks for single stars, and the
spectroscopy of these objects confirms that they are extremely hot
(see Figure 3).  U1 and U2 are consistent with blackbody temperatures
of 400,000~K and 500,000~K, respectively, although both the UV
photometric color and far-UV spectroscopic slope are becoming degenerate
with temperature at such temperatures.  More physically plausible
would be the synthetic spectrum of a hot star.  The hottest model in
the TheoSSA database of synthetic spectra for hot compact stars has a
temperature of 250,000~K, and is in reasonable agreement with the UV
spectral slope and photometry (Figure 3).  However, if U1 and U2 have
effective temperatures near 250,000~K, neither is actually compact;
stars with that temperature and with the far-UV luminosities we observe would
have radii similar to that of a low-mass post-AGB star that is about
to descend the white dwarf cooling curve.  The luminosities of U1 and
U2 would be log $L / L_\odot$~=3.1 and 3.5, respectively; such a high
luminosity is at the extreme limit of that found for any globular
cluster post-AGB star, and such temperatures are a factor of two
higher than those found in low-mass post-AGB tracks.  Note that some
estimates for the foreground extinction toward NGC~2808 are higher
than our value of $E(B-V) = 0.18$~mag.  For example, Harris (1996)
obtains $E(B-V) = 0.22$~mag from his assessment of the literature.
Assuming a higher extinction would imply that even hotter intrinsic
temperatures for U1 and U2 are needed to match the observed photometry
and spectroscopy.  Another possibility is that the extinction along
the NGC~2808 sightline differs from the Galactic mean curve. Neither
U1 or U2 appear to be coincident with x-ray sources in NGC~2808
(Servillat et al.\ 2008), but perhaps these objects are associated
with accretion disks that could explain their unusually hot
temperatures.

\begin{figure*}[t]
\plotone{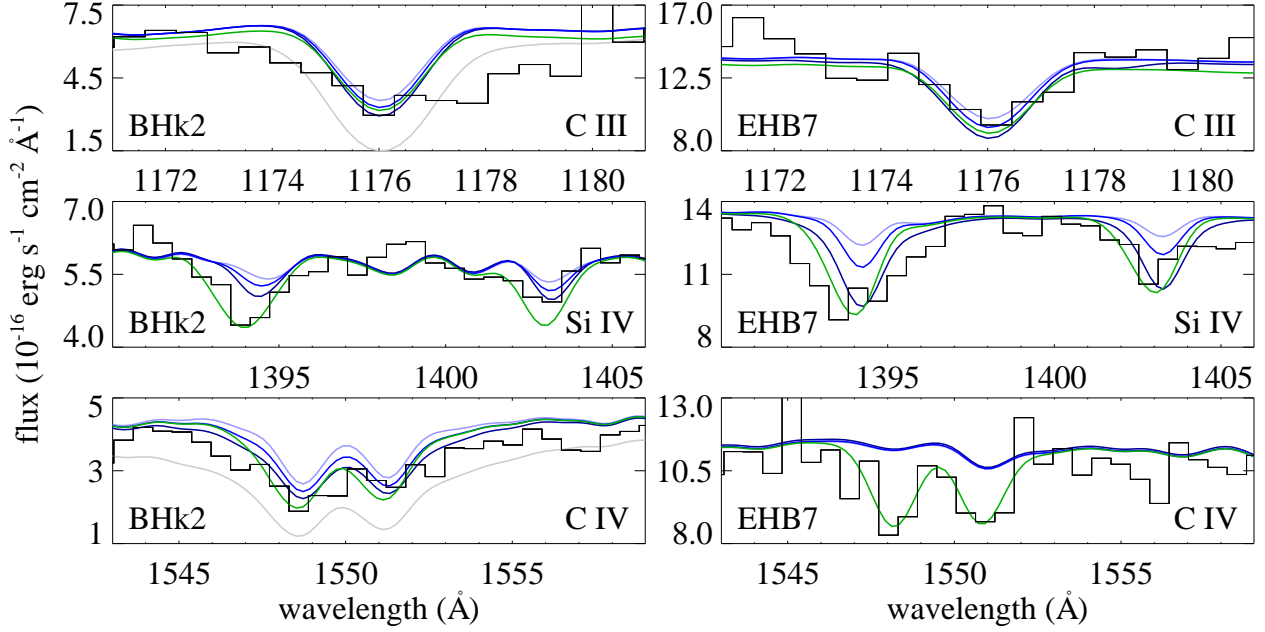}
\caption{
{\it Left panels:}
The spectrum of BHk2 (black histogram), with the same model shown in
Figure 6 (without ISM absorption, blue; with ISM absorption, green),
in the vicinity of strong C and Si features.  The resolution and SNR
of the spectra in this program are sufficient to detect the gross
abundance changes expected from flash mixing, but the uncertainties in
our abundances are large.  For comparison, we show two additional
models (without ISM absorption): one with the Si reduced by a factor of
10 and the C reduced by a factor of 2 (lighter blue), and one with the
Si increased by a factor of 10 and the C increased by a factor of 2
(darker blue).  The \ion{C}{3} feature (top panel) has almost no
interstellar contribution, but the feature is not particularly
sensitive to abundance in this regime, and this region of the spectrum
is adversely affected by the proximity to the Lyman-$\alpha$
geocoronal line.  The \ion{C}{4} feature (bottom panel) gives a good
indication of the C abundance, but is complicated by the 
contribution from interstellar absorption, which can be significant
for stars at at low $T_{\rm eff}$ and low C abundance (see also Figures 3--10).  
The \ion{Si}{4} feature (middle panel) has a large contribution 
from interstellar absorption, and the photospheric component varies weakly
with abundance.  Although the C abundance in the BHk stars is strongly
enhanced with respect to the normal EHB stars, it is not as strong as
the 3\% by mass one might expect for a star that has recently emerged
from flash mixing.  Despite the large uncertainties in the C
abundance, a 3\% C by mass can be firmly ruled out (grey).
{\it Right panels:} The same features, but in the spectrum of EHB7, one of our
coolest EHB stars.  The model from Figure 6 is shown for comparison
(without ISM absorption, blue; with ISM absorption, green), along with
two additional models (without ISM absorption) that have C and Si
reduced (lighter blue) and increased (darker blue) by factors of 2 and
10, respectively.  For stars at low C abundance and the low end of the
$T_{\rm eff}$ range in our sample (such as EHB7), the \ion{C}{4}
feature is dominated by ISM absorption, and so the C abundance must be 
determined from the \ion{C}{3} feature.  
}
\end{figure*}

The unclassified star U3 is very puzzling.  It has a very blue
$m_{FUV}-m_{NUV}$ color in the NGC~2808 photometry, but its far-UV spectrum is
much flatter than one would expect from a star with these UV colors.
Perhaps the photometry and spectrum are the result of some kind of blend,
non-stellar source, and/or circumstellar extinction, but we are unable
to put forth a plausible explanation for the object.
Like U1 and U2, U3 is not
coincident with x-ray sources in NGC~2808 (Servillat et al.\ 2008).
U2 and U3 respectively define the blue and red ends of a curious
string of stars bluer than the canonical HB but all sharing
approximately the same luminosity (see Figure 2).

\begin{figure*}[t]
\plotone{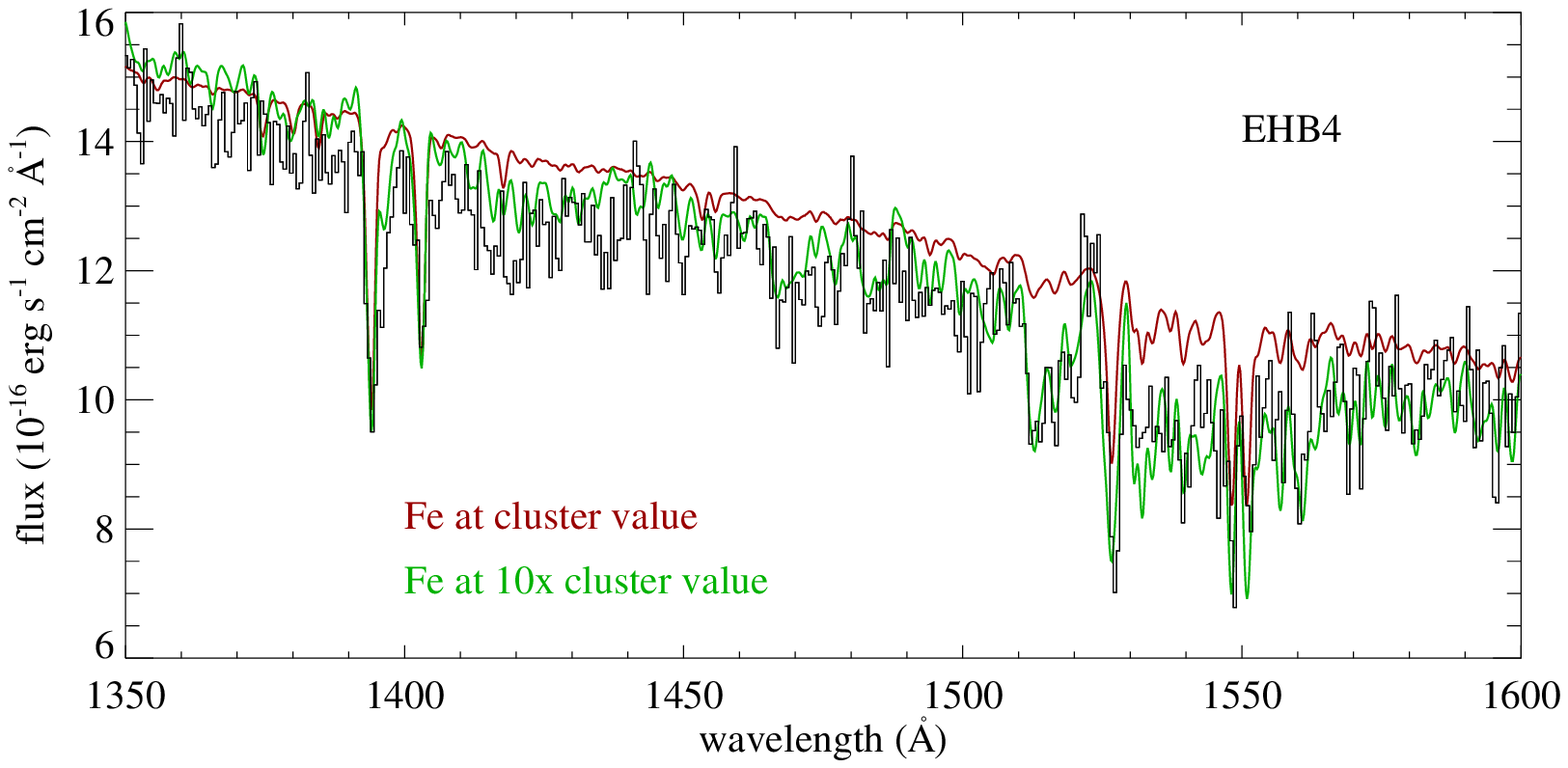}
\caption{
The spectrum of EHB4 (black histogram) in the vicinity of strong 
line-blanketing from
the Fe-peak elements.  For comparison, we show the best-fit model,
where these elements are enhanced with respect to the cluster value by
a factor of 10 (green; Figure 5), and a model where these elements
have not been enhanced (dark red).  We show an order-of-magnitude
variation for clarity, given that the abundance estimate comes from a
series of features over a broad wavelength range, but our uncertainty
in the abundance of the Fe-peak elements is approximately a factor of 3.
}
\end{figure*}

\subsection{BHB and AGBM Spectra}

The AGBM and BHB stars in Figures 9 and 10 are well-matched by spectra
from the UVBLUE (Rodr\'iguez-Merino et al.\ 2005) grid,
once these are interpolated to the NGC~2808 mean metallicity of
[Fe/H]=$-1.36$ and to the effective temperatures that match the
$m_{FUV}-m_{NUV}$ colors.  The UVBLUE spectra do not provide the
ability to independently vary individual elements, but the comparison
between the synthetic spectra and data shows no gross signatures of
atmospheric diffusion.  The equivalent widths of the C and Si lines in
our observed spectra are approximately matched by the lines in the
synthetic spectra, although the C and Si lines are mostly dominated by
interstellar absorption at these cooler temperatures.  Furthermore,
our AGBM and BHB spectra do not exhibit broad absorption troughs from
the Fe-peak elements, indicating that the abundances of the these
elements are not greatly enhanced.  This was unexpected, because our
BHB sample brackets the temperature ($\sim$11,500~K) where a discontinuity
occurs in the Str\"omgren photometry of several globular clusters
(Grundahl et al.\ 1999), which can be understood via sophisticated models
of stellar evolution that self-consistently include the effects of
atmospheric diffusion in the presence of turbulence (Michaud et al.\ 2008).
In the globular cluster M13, which has a
metallicity of [Fe/H]~=$-1.5$ (i.e., only slightly lower than that of
NGC~2808), the BHB stars hotter than this temperature exhibit an
enhancement in the Fe abundance that approaches three times the solar
abundance (Behr et al.\ 1999) -- one to two orders of magnitude higher
than cooler BHB stars and the mean cluster abundance.  In NGC~2808
itself, Pace et al.\ (2006) found that BHB stars at or below 12,000~K
exhibited no Fe enhancement, but that [Fe/H] increases to
$\sim$$-0.7$ at 12,200~K, $\sim$0.1 at temperatures of
12,400--12,800~K, and then $\sim$0.5--1.0 for stars hotter than
13,000~K.  Assuming the same relationship between Fe enhancement and
$T_{\rm eff}$ applies in our own sample, we would expect no
enhancement in BHB3, BHB4, and BHB5, but enhancements of one to two
orders of magnitude in BHB1 and BHB2.  Such enhancements would be very
obvious in the BHB1 and BHB2 spectra, but it is clear from Figures 9 and
10 that there is, in fact, little distinction in Fe abundance between
BHB2 and BHB3 (the two stars bracketing the temperature of this transition).

Because we do not observe a jump in Fe abundance within the
temperature range of our BHB sample, we explored the possibility that
our temperatures are systematically too hot, but this does not seem
likely.  For example, by assuming a significantly lower extinction of
$E(B-V) = 0.13$~mag, one could adopt temperatures for the BHB stars
that are $\sim$1,000~K cooler, thus placing them all below the
temperature of 12,000~K where one expects a jump in the Fe abundance
(see grey curves in Figure 9).  However, the resulting far-UV
synthetic spectrum has a Lyman-$\alpha$ profile much wider than that
observed, and the extinction would be much lower than most values
adopted in the literature, including that of Pace et al.\ (2006), who
assumed $E(B-V) = 0.22$~mag in their analysis.  If one does not resort
to a lower extinction, both the UV color and Lyman-$\alpha$ profile
place tight constraints on the temperatures of the BHB stars.  The
statistical error in UV color is less than 0.02~mag for each star (see
Figure 2), and so adopting a lower $T_{\rm eff}$ for each BHB star
would require significant systematic errors in the STIS photometric
calibration and also significant systematic errors in the
Lyman-$\alpha$ profile of each model, both going in the same
direction.  In BHB1, adopting a $T_{\rm eff}$ of 12,200~K instead of
12,900~K at our nominal $E(B-V) = 0.18$~mag would incur a mismatch in
UV color of 0.05~mag, with a gross mismatch in Lyman-$\alpha$ profile.
Because the temperature leverage provided by the UV color improves at
cooler temperatures, the constraints for the cooler BHB stars are even
more stringent.  Specifically, if one is willing to accept a 0.05~mag
mismatch in UV color, BHB2 could be 600~K cooler, BHB3 could be 450~K
cooler, BHB4 could be 200~K cooler, and BHB5 could be 150~K cooler.
In each case, however, the Lyman-$\alpha$ profile of the cooler model
would be much broader than observed.

The star AGBM1 is far brighter than the other stars in our sample, and
its spectrum has the highest SNR.  Although the observed spectrum is
generally well-matched by the interpolated UVBLUE spectrum, there
is a discrepancy in the strength of the
interstellar \ion{Si}{2} feature at 1190~\AA\ (which arises from our
interstellar absorption model, and not the UVBLUE grid).  Although
this feature falls within the region where the SNR is depressed by the
broad geocoronal Lyman-$\alpha$ line, the luminosity of this
particular star means it still has adequate SNR to accurately measure
this feature if present.  In fact, the \ion{Si}{2}
$\lambda$1190~\AA\ feature is overpredicted by our interstellar
absorption model for all of the stars in our entire sample, but our
chosen \ion{Si}{2} column does reproduce the other interstellar
\ion{Si}{2} features in the spectra.

\section{Discussion}

Compared to the stars in our EHB sample, the stars in our BHk sample
are significantly hotter, and have much higher He and C abundances.  This
can be seen from both an empirical comparison of the mean spectrum for
each class, and also from a comparison of the models that best
reproduce the individual stars in each class.  NGC~2808 likely hosts a
sub-population of MS stars born with an enhanced He abundance
($Y\sim0.4$; D'Antona et al.\ 2005; Piotto et al.\ 2007), and such
stars are more likely to produce EHB stars (including the EHB stars of
normal luminosity and the subluminous BHk stars).  However, the
flash-mixing scenario is the only plausible mechanism for producing the
higher temperatures and enhanced He and C abundances in our BHk sample.

\begin{figure*}[t]
\plotone{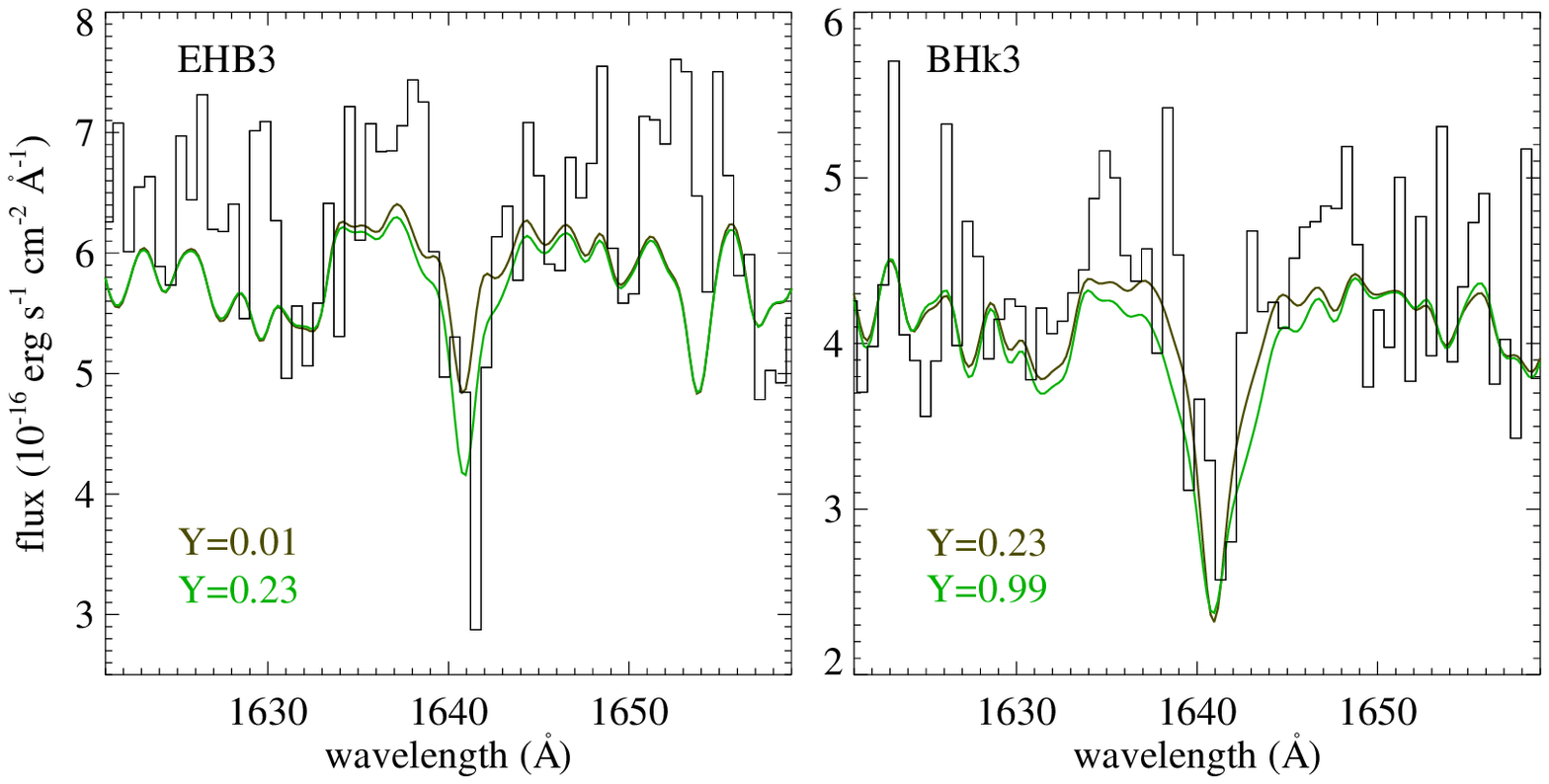}
\caption{{\it Left panel:}
The spectrum of EHB3 (black histogram), shown with the $Y=0.23$ model of 
Figure 4 (green), compared to a model with all of the same parameters except for
He abundance, which has been reduced to $Y=0.01$ (brown).  
EHB stars are frequently depleted in He (e.g., Moehler et al.\ 2011), but
in our sample, the spectra of EHB1 and EHB3 exhibit a detectable
\ion{He}{2} 1640~\AA\ feature.  The equivalent width of this feature in the data
is better matched by the feature in the $Y=0.23$ model
than that in the $Y=0.01$ model, although
the feature is difficult to characterize when it is weak, given the line
blanketing in this region and the relatively low SNR.  Lowering the He
from $Y=0.23$ to $Y=0.01$ has a negligible effect on the rest of the
far-UV spectrum and the $m_{FUV}-m_{NUV}$ color; the $Y=0.01$ model would
be difficult to distinguish if shown in Figure 4.  Aside from EHB1 and EHB3,
the remaining EHB stars in our sample could be significantly depleted in He, 
but because they are significantly cooler, this feature is too weak to provide
meaningful constraints on the He abundance.
{\it Right panel:} The spectrum of BHk3, shown with the $Y=0.99$ model of 
Figure 7 (green), compared to a model with all of the same parameters
except for He abundance, which has been reduced to $Y=0.23$ (brown).
The \ion{He}{2} 1640~\AA\ feature is better-matched by the $Y=0.99$ model 
(green), but, more importantly, the $Y=0.23$ model has a much
bluer $m_{FUV}-m_{NUV}$ color ($-$1.60~mag) than the $Y=0.99$ model 
($-$1.48~mag). If one reduces the $T_{\rm eff}$ in the $Y=0.23$ model to try 
to match the $m_{FUV}-m_{NUV}$ color in the data ($-$1.47~mag), 
the \ion{He}{2} feature in the model becomes even weaker, increasing the 
discrepancy with the feature observed in the spectrum.  Moreover,
the model does not match the observed far-UV spectral slope.
Thus the simultaneous fitting of the \ion{He}{2} feature, the 
$m_{FUV} - m_{NUV}$ color, and the far-UV spectral slope points to a high He
abundance for BHk2.
}
\end{figure*}

Although the C abundance in our BHk sample is greatly enhanced relative
to the EHB sample, it is not as high as one would expect for stars
that underwent flash mixing very recently.  The strongest enhancements
of C in our BHk sample are at 10$\times$ the cluster ratio of C to H,
corresponding to a mass fraction of 0.2\% in the atmosphere, whereas
flash mixing is expected to produce stars with atmospheric C
abundances of 1--4\% by mass.  The reduction in C abundance is not
unexpected, however.  Miller Bertolami et al.\ (2008) calculated
numerical simulations of stars as they evolved through the
flash-mixing stage and subsequent period of stable core He burning,
taking into account the atmospheric diffusion processes of
gravitational settling and radiative levitation.  They found that
stars emerge from the flash-mixing process with strongly enhanced He
and C abundances, but that the C abundance declines rapidly, dropping
by an order of magnitude after 1000 yr, and by several orders of
magnitude by the time the star has evolved for some
10$^7$ yr after the flash mixing of the envelope.  Thus, the star
spends only a very small fraction of its stable core He-burning lifetime
with a C abundance near its maximum of 1--4\% by mass.  Their calculations
also show the He will eventually decline in the atmosphere as well,
although this happens over a much longer timescale, such that the 
decline does not become significant until $\sim$10$^6$--10$^7$ yr later.
The fact that Lanz et al.\ (2004) found higher C abundances in two of 
the three He-sdB stars in their Galactic field sample may be a selection
effect; in NGC~2808, the stars were selected by position in the UV
CMD, but in the Galactic field, the stars were selected by He abundance.
Even in globular clusters, there might be other systematic effects at play.  
In their optical spectroscopy of candidate BHk stars in 
$\omega$ Cen, Moehler et al.\ (2011) found significant enhancements
of He and C in the hottest HB stars, but in their sample the C enhancement
approached the 3\% one would expect for stars that have recently undergone
flash mixing.  Whether this is due to systematic differences in approach
(e.g., optical vs. UV spectroscopy) or intrinsic differences in the 
populations is unclear.

The variations in Si and Fe-peak abundances clearly
demonstrate that atmospheric diffusion is significant in the EHB and
BHk populations.  Abundance anomalies have also been
well-documented in the field population of sdB stars (e.g., 
Heber et al.\ 2000; O'Toole \& Heber 2006; Blanchette et al.\ 2008; 
Heber 2009; Geier et al.\ 2010), where large enhancements in the Fe-peak
elements (but not Fe itself) are observed.  However, recent calculations 
demonstrate that other processes, such as turbulent mixing and mass loss, 
may also play a role in these anomalies (Hu et al.\ 2011; 
Michaud et al.\ 2011).  Our observations
demonstrate that large enhancements in the Fe-peak elements can occur
in the BHk stars, and may explain those BHk stars with
unusually red UV colors (Brown et al.\ 2001, 2010).  Two of the BHk
stars in our spectroscopic sample were drawn from this unusually red
segment of the BHk population in NGC~2808, and both exhibit enormous
enhancements in the Fe-peak elements (25--100$\times$ the cluster
value).  Apparently a large dispersion in the abundances of Fe-peak
elements, when combined with flash mixing, can provide the large color
range observed in the BHk population of massive clusters.

The hottest BHk stars in our sample fall at temperatures similar to
those of recently-discovered pulsating subdwarfs in $\omega$ Cen
(Randall et al.\ 2011).  The four pulsating subdwarfs in $\omega$ Cen
are located near the BHk region of the $\omega$ Cen CMD, and Randall et
al.\ (2011) derive 48,000~K~$\lesssim T_{\rm eff} \lesssim$~52,000~K
from their optical spectroscopy.  Randall et al.\ (2011) suggest that
their sdO pulsators inhabit a newly-discovered
instability strip; if that is the case, our hottest BHk stars (BHk1 \&
BHk4) may also be pulsators.  The enhanced Fe-peak abundances observed
in our hottest BHk stars may be significant, as radiative levitation
appears to play a role in sdO pulsations (Fontaine et al.\ 2008).  We
note, however, that Randall et al.\ (2011) found subsolar He
abundances for their pulsators, in contrast to the He-rich atmospheres
found in our hottest BHk stars.  One intriguing possibility is that the
pulsators in $\omega$ Cen are evolved from BHk stars, which are known
to exist in $\omega$ Cen; diffusion processes can convert a He-rich BHk
star into a He-poor sdO (Miller Bertolami et al.\ 2008).  If flash
mixing is required to achieve these high EHB temperatures, then the
dearth of analogous pulsators in the field population may be
explained because flash mixing is more likely to occur in populations born at
high He abundance ($Y \sim 0.4$), such as those subpopulations found
in massive globular clusters.

\acknowledgements

Support for Program 11665 was provided by NASA through a grant
from STScI, which is operated by AURA, Inc., under NASA contract NAS
5-26555.  The TheoSSA service (http://dc.g-vo.org/theossa) used to retrieve
a theoretical spectrum for this paper was constructed as part of the
activities of the German Astrophysical Virtual Observatory.  We thank 
the anonymous referee for useful suggestions that improved the clarity
of our manuscript.

\end{document}